\documentclass[useAMS,usenatbib]{mnras}

\usepackage[T1]{fontenc}
\usepackage{ae,aecompl}
\usepackage{array}

\usepackage{graphicx}
\graphicspath{{figures/}}	
\usepackage{amsmath}	
\usepackage{epstopdf}
\usepackage{multirow}
\usepackage{amssymb}

\usepackage{newtxtext,newtxmath}
\usepackage[normalem]{ulem} 
\usepackage{makecell} 

\title[100\,pc samples and SFH]{Assessing the star formation history of all-sky and part-sky 100\,pc white dwarf samples}

\author[Roberts et al.]{Emily K. Roberts$^1$\thanks{E-mail: emilyroberts0311@gmail.com}, 
Pier-Emmanuel Tremblay$^1$ and Antoine Bédard$^1$
  \\
$^{1}$Department of Physics, University of Warwick, Coventry CV4 7AL, UK 
}

\date{Accepted XXX. Received YYY; in original form ZZZ}

\pubyear{2025}

\begin{document}
\label{firstpage}
\pagerange{\pageref{firstpage}--\pageref{lastpage}}
\maketitle

\begin{abstract}
Thanks to \textit{Gaia}, as well as present and future data from large-scale multi-object spectroscopic follow-up surveys (4MOST, DESI, WEAVE, SDSS-V), it is now possible to build representative and minimally biased samples of the local white dwarf population.
Here we analyse several volume-limited 100\,pc samples of white dwarfs, constructed from different surveys and studies, to evaluate their completeness and residual biases. We model the underlying star formation history and Galactic disc age via comparison with simulated populations of white dwarfs to quantitatively characterise completeness. We assess whether the benefit of \textit{Gaia} XP spectra in datasets outweighs the reduction in sample size, and to what extent targeted, part-sky, and magnitude limited surveys can be used in comparison to all-sky volume limited surveys. Additionally, we simulate the 4MOST 100PC spectroscopic sub-survey and discuss its use to better understand the local star formation history.
\end{abstract}

\begin{keywords}
white dwarfs -- stars: luminosity function -- stars: evolution -- Galaxy: solar neighbourhood
\end{keywords}



\section{Introduction}\label{sec:intro}

White dwarfs have long been recognised for their potential as chronometers, since they cool over time and have no source of nuclear fuel to maintain their temperatures \citep{Mestel}. With a close relationship between effective temperature, luminosity (or mass), and their total age (main-sequence lifetime plus white dwarf cooling time), they provide a way of rewinding time in a stellar population to look at the underlying star formation rates \citep{Winget_1987,Oswalt_1996,Rowell_2013,Tremblay_2014,Isern_2019,Fantin_2019,Torres_2021,Cukanovaite_2023,Roberts_2025}. Realising this potential, however, requires detailed models of white dwarf cooling times, main sequence lifetimes, the initial mass function, the initial-final mass relation, and a careful treatment of observational biases. These variables contribute uncertainty and cloud our knowledge of the star formation history (SFH) that produced the white dwarfs. 
A common benchmark is to compare such reconstructions with the formation histories inferred from main-sequence stars in the same populations \citep{Mor_2019,Alzate_2021,DalTio_2021,Gallart_2024,Gallart_2025}, with the broader goal of informing models of Galactic formation and evolution.


Several methods have been used to derive the SFH of a white dwarf population based on different observational properties: estimating individual total ages for each star or using stellar population synthesis models to fit the luminosity function, absolute magnitude distribution, or Hertzsprung–Russell diagram. In a recent study, \citet{Roberts_2025} compared these approaches using the 40\,pc \textit{Gaia}-identified, volume-complete sample of 1073 spectroscopically confirmed white dwarfs \citep{OBrien_2024}. They found that, within systematic uncertainties and finite sample size, the different methods yielded broadly consistent results. Among them, the method based on the absolute \textit{Gaia G}-band magnitude distribution was identified as the most promising, since it requires only a parallax and a \textit{G}-band magnitude and can therefore be applied to larger and more distant samples with less complete follow-up data. Using these methods, they showed that a constant SFH provided a good overall fit to the sample, although variations by up to a factor of about two — either in the recent past \citep[e.g.][]{Mor_2019} or at earlier times—could not be ruled out given the uncertainties in population modelling. For example, the SFHs of \cite{Mor_2019} and \cite{Roberts_2025} from direct age method have opposite behaviour at old times but are statistically equivalent fits to the 40\,pc sample \citep{Roberts_2025}.





With each successive \textit{Gaia} data release \citep{Gaia_2016b,Gaia_2018,GaiaDR3_2023}, the size and precision of stellar samples available for Galactic studies has grown dramatically. The advent of machine learning classification, improved stellar models and quality filtering, and increasingly realistic Galactic simulations have further enhanced our ability to extract astrophysical information from these data. In parallel, large efforts have been devoted to the spectroscopic characterisation of \textit{Gaia} catalogues: from the low-resolution XP spectrophotometry provided on board \textit{Gaia}, to ground-based multi-object spectroscopic (MOS) surveys such as DESI \citep{DESI_2025}, 4MOST \citep{4MOST}, WEAVE \citep{WEAVE}, and SDSS-V \citep{SDSS-V}, which together provide the medium-resolution ($R \approx 5000$) optical ($\approx 3600-10\,000$\,\AA) spectroscopic context essential for Galactic archaeology and beyond. This enables more precise determinations of stellar parameters, chemical abundances, radial velocities, magnetic field strengths, and binary properties that are not possible with \textit{Gaia} alone. 

There now exist a range of 100\,pc \textit{Gaia}-defined samples of confirmed and candidate white dwarfs \citep{GentileFusillo_2021,GCNS_2021,JimenezEsteban_2023,Garcia-Zamora_2023,Vincent_2024}, constructed with varying quality cuts, white dwarf probability thresholds, and each of them can be combined with auxiliary data, leading to differences in sky coverage. In particular, the footprint of \textit{Gaia}-selected white dwarfs observed in MOS surveys currently covers only about 30\% of the sky (DESI DR1; \citealt{DESI_2025}, SDSS; \citealt{Kleinman2013}, LAMOST; \citealt{Lamost2012}), with plans for each MOS mission to expand to roughly 50\% of the sky by the end of operations. While these catalogues contain together and within 100\,pc an order of magnitude more white dwarfs than the 40\,pc sample, reducing Poisson errors on the SFH, and allowing for a wider probe of the Solar neighbourhood including the effect of the finite vertical scale height of the Galactic disc, they also lack certain things the 40\,pc sample has such as either complete spectroscopic follow up or full volume completeness. Balancing these factors and assessing their impact is crucial for selecting the best white dwarf samples for population studies.

In this work, based on the 100\,pc \textit{Gaia} stellar sample, we compare different white dwarf sample selections, either based on quality cuts or availability of auxiliary data such as XP spectrophotometry or MOS medium-resolution spectroscopy. We then rely on the stellar population code of \cite{Roberts_2025} to quantify these differences in terms of changes in inferred SFH and age of the Galactic disc.



In Section \ref{sec:100pc}, we outline the different 100\,pc \textit{Gaia} samples we work with. Section \ref{sec:methods} recaps the three methods of determining the SFH of a population, while Section \ref{sec:sim} explores the uncertainties in our modelling and their impact. In Section \ref{sec:results} we show the results of our simulations for the 100\,pc samples and examine their implications. Finally, in Section \ref{sec:conc} we discuss the conclusions of this work and its prospects for future surveys.

\section{100\,pc white dwarf samples}\label{sec:100pc}

Identifying white dwarfs in the \textit{Gaia} DR3 catalogue is challenging, even within the relatively nearby 100\,pc volume. As shown for example by \citet{GentileFusillo_2021}, contamination from non-white dwarf objects remains non-negligible, requiring probabilistic classification to separate white dwarfs from main-sequence stars, unresolved binaries and spurious sources. Nevertheless, 
the completeness of the \textit{Gaia} 100\,pc white dwarf sample is expected to be high, on the order of 93 per cent from a comparison with the SDSS footprint \citep{GentileFusillo_2021}. 
This is in line with an independent estimate by \citet{OBrien_2023,OBrien_2024}, which suggests that the \textit{Gaia} white dwarf selection of \citet{GentileFusillo_2021} achieves $\approx$97 per cent completeness in the  40\,pc sample from a comparison with previously known white dwarfs not identified in \textit{Gaia}, many of them members of unresolved binaries. 


The sensitivity of the 100\,pc selection is also demonstrated by different authors reaching different selections. The \textit{Gaia} Collaboration's Catalogue of Nearby Stars \citep{GCNS_2021} (GCNS) finds 14\,699 high-probability white dwarfs ($P_{\rm WD}\geq0.75$) within 100\,pc compared to the 15\,001 found in \citet{GentileFusillo_2021} for the same threshold in $P_{\rm WD}$. Only 14\,687 of these sources are in common between both catalogues. Also, the independent analysis of \citet{JimenezEsteban_2023} arrives at a different total of 12\,718 white dwarfs within 100\,pc. These reflect the different HR diagram selections, definition of white dwarf probability thresholds, and adopted quality cuts, underscoring the difficulty of assembling a complete volume census from \textit{Gaia} alone.

Restricting the population to those white dwarfs with additional spectroscopic information reduces the available sample substantially. Limiting to objects with \textit{Gaia} XP spectrophotometry, as explored by \citet{Vincent_2024}, results in 13\,235 white dwarfs within 100\,pc. Limiting to objects with existing ground-based spectroscopy within the SDSS footprint \citep{Kilic_2025_100pc} or DESI DR1 \citep{DESI_2025} results in samples of 2\,750 and 2\,978 white dwarfs, respectively.  While these cuts enhance the reliability of atmospheric parameter estimates and classifications, they inevitably reduce the sample size and may introduce biases. Most of this loss arises from the limited (primarily northern hemisphere) footprints of these MOS surveys, but within their surveyed conical 100\,pc volumes the resulting samples remain reasonably complete and representative \citep{Kilic_2020,Manser_2024,Kilic_2025_100pc}. 

This work aims at directly comparing these different samples, summarised in Table~\ref{tab:numbers_in_100pc}, and quantifying their differences regarding overall white dwarf population properties and inferred SFH. Furthermore, we aim at qualifying the prospects of ongoing MOS surveys with a 100\,pc footprint (DESI, 4MOST, WEAVE, SDSS-V) in lessening the biases of current samples.

\renewcommand{\arraystretch}{1.8}

\begin{table*}
	\centering
	\caption{Sample sizes and quality cuts made for the samples considered in this work. The part-sky DESI and SDSS footprints are shown in Fig.\,\ref{fig:footprints}. In all cases, photometric atmospheric parameters were corrected for the low-mass problem at $T_{\rm eff}<6000$\,K \citep{OBrien_2024} before the mass cuts for binary removal were performed (see ``After mass cuts" column and text in Section~\ref{sec:100pc}). Note that in the absolute $G$ magnitude distribution method, we do not otherwise use the derived stellar parameters. The Jim{\'e}nez-Esteban sample does not have masses and effective temperatures for all stars in the sample; where they are missing, the values from \citet{GentileFusillo_2021} are used instead. `tailored atm' implies that photometric or spectroscopic fits were performed with model atmospheres using an appropriate chemical composition for a given known spectral type. The GCNS catalogue \citep{GCNS_2021} also includes main-sequence stars, hence the large number of objects removed after the $P_{\rm WD}$ white dwarf selection. The column ``Overlap with GF21" gives the overlap with the reduced 100\,pc sample of \citet[][first row]{GentileFusillo_2021} containing 15\,001 white dwarfs for reference. All samples except GCNS are subsets of the full catalogue of \citet{GentileFusillo_2021} given the $P_{\rm WD}$ cut, but the final membership depends on the additional mass cuts (see footnote 1).}
	\label{tab:numbers_in_100pc}
	\begin{tabular}{||lllllll||} 
 \hline
  Source          & Type    & \thead[l]{Stars within\\  100\,pc in sample} & \thead[l]{After $P_{\rm WD}$\\$\ge$0.75 cut} & \thead[l]{After \\ mass cuts} & \thead[l]{Overlap\\with GF21} & $T_{\rm eff}$/$\log(g)$/mass source \\ \hline \hline
\thead[l]{\citet{GentileFusillo_2021}\\(GF21; \textit{Gaia} selection)} & \thead[l]{All-sky\\100\,pc} &         16298           &  16281        &          15001  &     15001     & \thead[l]{\citet{GentileFusillo_2021}\\photometric \textit{Gaia}, H-atm assumed}           \\ \hline
\thead[l]{\citet{GCNS_2021}\\(GCNS; \textit{Gaia} selection)}          & \thead[l]{All-sky\\100\,pc} &         301928          &  16319        &          14699  &     14687   & \thead[l]{\citet{GentileFusillo_2021}\\photometric \textit{Gaia}, H-atm assumed}           \\ \hline
\thead[l]{\citet[][]{Vincent_2024}\\(GSPC-WD \textit{Gaia} XP sample)}        & \thead[l]{All-sky\\100\,pc} &        14906           &  14802        &          13235  &     12889   & \thead[l]{\citet{Vincent_2024}\\photometric \textit{Gaia} XP, tailored atm}   \\ \hline
\thead[l]{\citet{Kilic_2025_100pc}\\(Spectro. SDSS sample)}           & \thead[l]{Part-sky\\100\,pc} &        3145            &  3137         &          2750   &     2691    & \thead[l]{\citet{Kilic_2025_100pc}\\photo. and spectro., tailored atm}  \\ \hline
\thead[l]{\citet{DESI_2025} + this work\\ (Spectro. DESI sample)}          & \thead[l]{Part-sky\\100\,pc} &        3240            &  3219         &          2978   &     2973    & \thead[l]{\citet{GentileFusillo_2021}\\photometric \textit{Gaia}, H-atm assumed}            \\ \hline
\thead[l]{\citet{JimenezEsteban_2023}\\(\textit{Gaia} selection)} & \thead[l]{All-sky\\100\,pc} &    12718                  &  12580       &        11118   &     11069   & \thead[l]{\citet{JimenezEsteban_2023}\\photometric \textit{Gaia} XP, tailored atm}  \\ \hline
\thead[l]{\citet{OBrien_2024}\\(\textit{Gaia} selection)}    & \thead[l]{All-sky\\40\,pc} &        1073            &   1073        &          960    &     956     & \thead[l]{\citet{GentileFusillo_2021}\\photometric \textit{Gaia}, tailored atm}  \\ \hline
 \end{tabular}
\end{table*}

\subsection{All-Sky 100\,pc Samples}


We consider three \textit{Gaia} defined all-sky 100\,pc white dwarf samples from Early Data Release 3 and Data Release 3. The first and second, \cite{GentileFusillo_2021} and \cite{GCNS_2021}, are selected based on \textit{Gaia} EDR3 photometry and astrometry, which remained unchanged in DR3. Finally, the sample of \cite{Vincent_2024} uses \textit{Gaia} DR3 and its new XP spectra to define an independent sample. While this sample is drawn from the full \textit{Gaia} DR3, rather than existing \textit{Gaia} EDR3 selections, its size and completeness are limited by the availability of \textit{Gaia} XP spectroscopy. For all samples we use \textit{Gaia G}, $G_{\rm BP}$, $G_{\rm RP}$, and parallax directly from \textit{Gaia} DR3.

\begin{figure}
    \includegraphics[width=0.9\columnwidth]{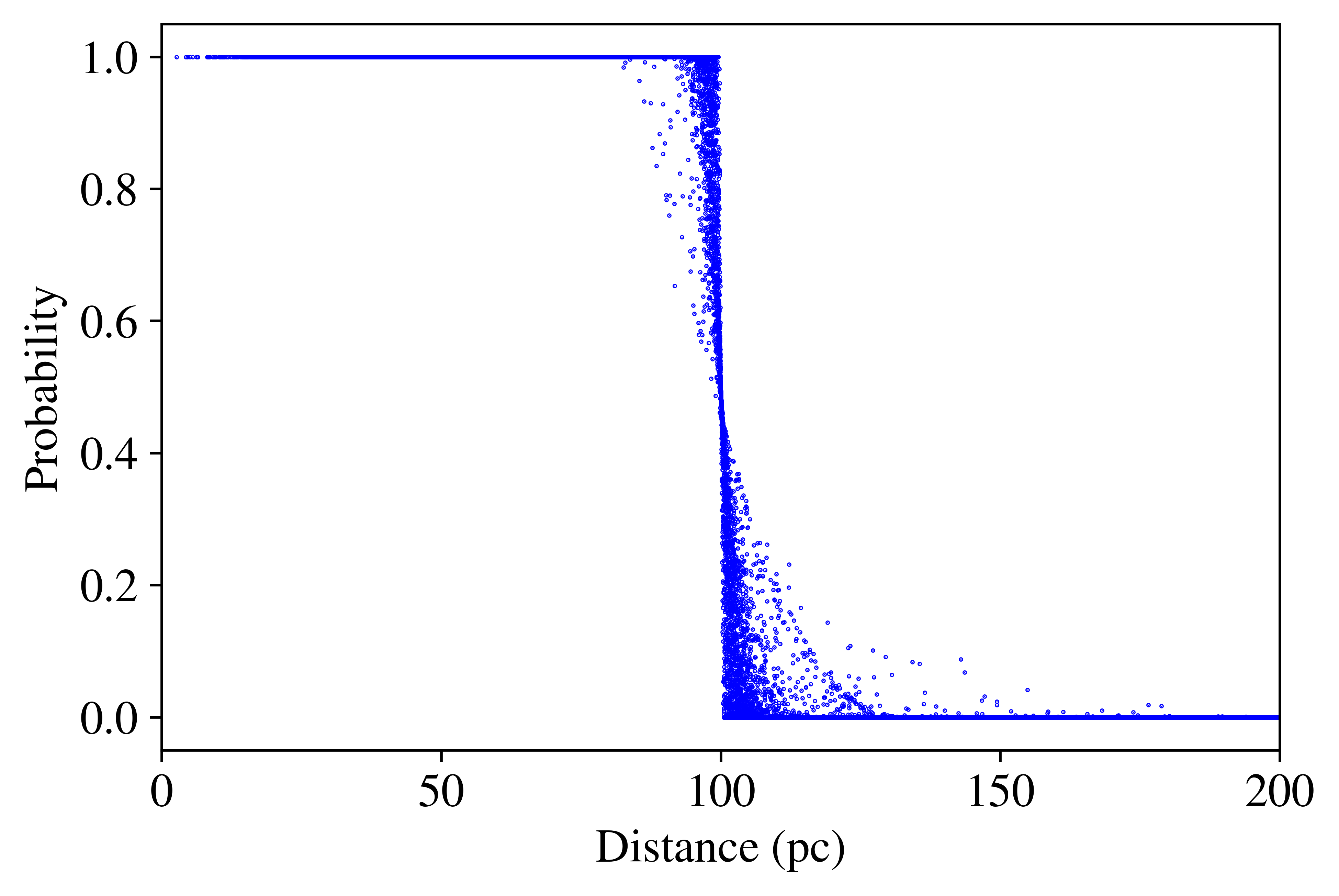}
    \caption{The probability of each white dwarf candidate in the \protect\citet{GentileFusillo_2021} sample lying within the 100\,pc volume based on its determined parallax and uncertainty on its parallax. 
    Fewer than 2\,000 objects outside the 100\,pc limit have any probability of being within 100\,pc within 1$\sigma$, and all of them have $P_{100 \rm pc}<0.5$.}
    \label{fig:prob_in_100pc}
\end{figure}

\textit{Gaia} provides parallax uncertainties, which allows for the calculation of the probability of any star being within 100\,pc, including those whose parallax values would put them outside that volume with strict cut offs, and also those nominally within 100\,pc that have a chance of being outside of the volume. We found that these probabilistic inclusions do not significantly change any of the results discussed in this paper, as for the vast majority of white dwarfs or white dwarf candidates, the probability of being in the volume is either $<1$ per cent or $>99$ per cent, with very few ambiguous objects as shown in Fig.\,\ref{fig:prob_in_100pc}. Fewer than 2\,000 objects outside the strict 100\,pc cut off have any probability at all of being within 100\,pc at 1$\sigma$, once parallax errors are considered, and all of them have a probability of being within 100\,pc of below 0.5. The majority of these possible 100\,pc candidates do have high $P_{\rm WD}$ but their inclusion in the 100\,pc sample does not change any of the results meaningfully therefore we do not include these possible 100\,pc candidates in any of the samples.

For the samples of \citet{GentileFusillo_2021} and \cite{GCNS_2021}, we adopt photometric masses and $T_{\rm eff}$ from \citet{GentileFusillo_2021}, assuming hydrogen atmospheres due to the lack of consistent composition data (spectroscopy) in these samples. For the \citet{Vincent_2024} sample, we adopt their photometric parameters based on \textit{Gaia} XP spectra and spectral classification, which enables the use of tailored model atmospheres for the corresponding spectral classes.

Due to inaccuracies in opacities in the red wing of Lyman\,$\alpha$ absorption \citep{Caron_2023,OBrien_2024,Sahu_2025}, the median photometric mass of white dwarfs as a function of temperature, which should remain at a constant $\approx0.6$\,M$_{\odot}$, begins to drop below $\approx6000$\,K in all samples here. To correct for this, we apply a fifth-order polynomial correction as defined by \cite{OBrien_2024} to bring the median masses back in line with $\approx0.6$\,M$_{\odot}$ for all samples discussed in this work, and before any sample cuts on masses. Since photometric temperature derivations are based on the assumed mass through the mass-radius relation and the Stefan-Boltzmann law, corrections to effective temperature are also made.\footnote{Since this correction uses the median mass of each sample, the exact mass and effective temperature correction applied to a star might not be the same depending on which sample it appears in. These differences are slight but in some cases cause stars to be removed from one sample but kept in another when we perform further cuts on masses, i.e. to remove unresolved double degenerates.} 

\begin{figure}
	\includegraphics[width=0.9\columnwidth]{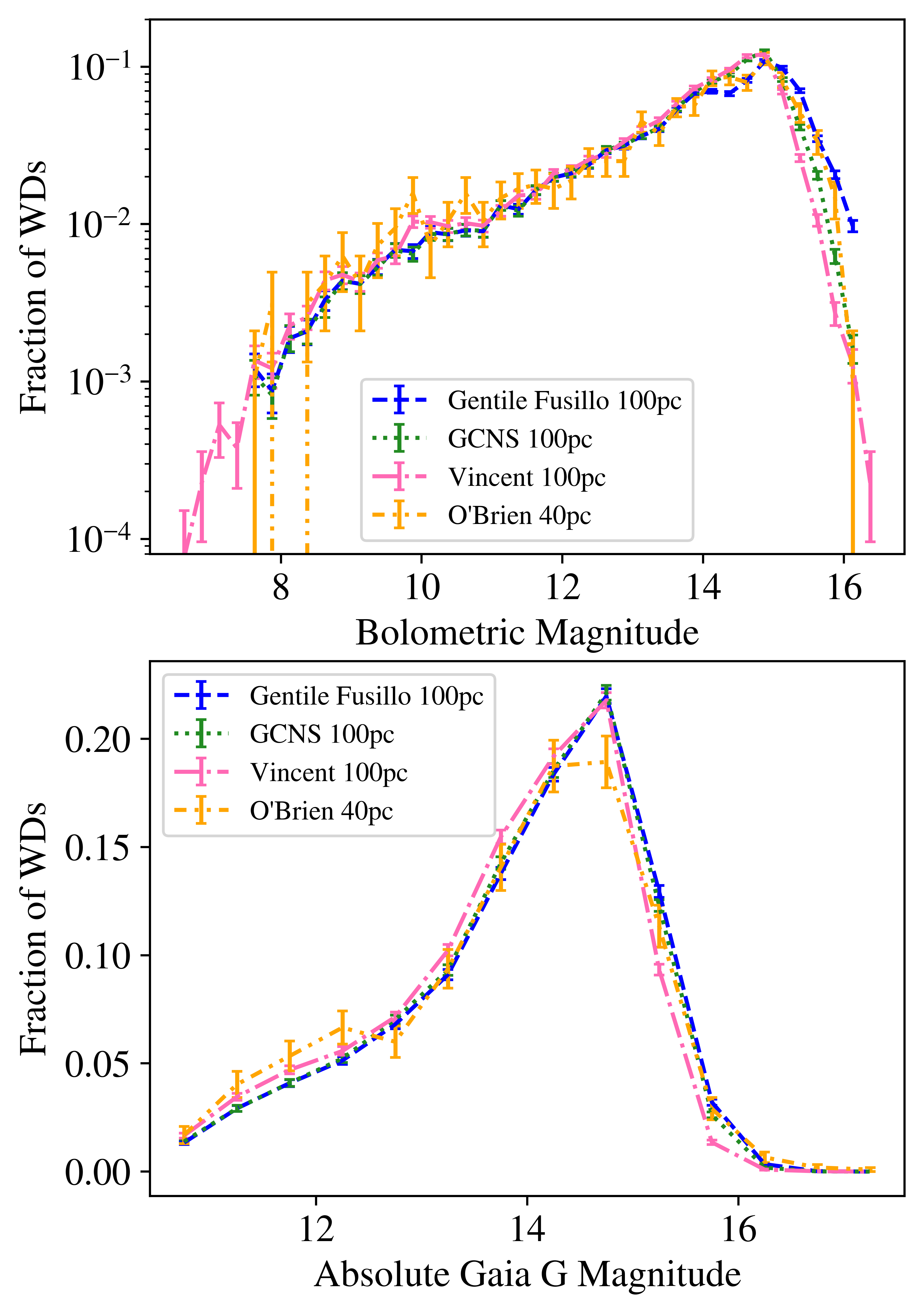}
    \caption{\textit{Top:} Luminosity function for each of the three all-sky 100\,pc samples of \citet{GentileFusillo_2021,GCNS_2021,Vincent_2024} and the 40\,pc sample \citep{OBrien_2024,Roberts_2025}. We also include the Poisson uncertainties for each bin.
    \textit{Bottom:} Absolute \textit{Gaia G} magnitude distribution for each of the  all-sky 100\,pc samples and the 40\,pc case, with Poisson uncertainties for each bin. The bump at $\approx$12.5 mag in the 40\,pc sample does not appear in any of the 100\,pc samples at the $3\sigma$ level and so appears to be an artifact of the smaller sample size rather than a real feature.}
    \label{fig:100vs40pc_results}
\end{figure}

The three all-sky samples are discussed in turn below. We also show their luminosity functions and absolute $G$ magnitude distributions in Fig.\,\ref{fig:100vs40pc_results}, compared to the 40\,pc sample of \citet{OBrien_2024,Roberts_2025}. The absolute $G$ magnitude distributions rely solely on sample membership and \textit{Gaia} DR3 data, while the luminosity functions rely on white dwarf parameters described in the last column of Table\,\ref{tab:numbers_in_100pc} along with the mass-radius relation of \citet{Bedard_2020}.

\subsubsection{Gentile Fusillo et al. sample}

With selection criteria in absolute magnitude, colour and \textit{Gaia} quality flags, \cite{GentileFusillo_2021} presents a catalogue of objects from \textit{Gaia} EDR3. They calculate the probability of each object being a white dwarf based on SDSS spectroscopically confirmed white dwarfs and contaminants to provide a catalogue of $\simeq$ 359\,000 high confidence white dwarf candidates ($P_{\rm WD} \ge 0.75$). Using these candidates, white dwarf masses, $T_{\rm eff}$ and $\log(g)$ were calculated using dereddened \textit{Gaia} photometry for pure hydrogen, pure helium and mixed hydrogen-helium atmosphere models. Because few candidates have spectroscopic confirmation, parameters based on all three model assumptions are provided in the catalogue, along with mapped $A_V$ extinction. 

Using the probability of being a white dwarf defined by \cite{GentileFusillo_2021} and keeping only those objects with $P_{\rm WD}\geq 0.75$, alongside a distance cut-off at 100\,pc, and limiting objects to those with a white dwarf mass between $0.54-1.34$\,M$_{\odot}$ to eliminate double degenerate candidates and physically implausible white dwarfs if born through single star evolution, the Gentile Fusillo et al. catalogue contains 15\,001 white dwarf candidates with corresponding physical parameters, which we use as a reference in Table\,\ref{tab:numbers_in_100pc} when calculating the overlap with other samples. While many factors such as metallicity, rotation and magnetic effects may produce lower mass white dwarfs through single star evolution \citep{Renedo_2010, Cunningham_2020}, these are not currently fully included in empirical initial-final mass relations. As an approximation, we therefore use the hard cut off of 0.54\,$M_{\odot}$. The upper limit is to enforce the Chandrasekhar mass limit for white dwarfs \citep{Chandrasekhar_1931}.

We note that unresolved white dwarf–main sequence binaries are already excluded to a large extent in the initial Hertzsprung-Russell diagram selection of \citet{GentileFusillo_2021}. Our additional double-degenerate cut is intended to define a sample primarily composed of present-day single white dwarfs, while possible past stellar mergers are accounted for in the population synthesis \citep{Roberts_2025}. Although removing unresolved binaries reduces the overall completeness of the sample, the resulting biases should remain small to moderate, since binary white dwarfs are expected to share a similar age distribution with single white dwarfs, the main parameter constrained in this work. Moreover, our analysis focuses on relative rather than absolute star formation rates.

\subsubsection{Gaia Catalogue of Nearby Stars sample}

The GCNS is a \textit{Gaia} Early Data Release 3 catalogue laid out in \cite{GCNS_2021}. They define the Solar neighbourhood to be all objects within a sphere of radius 100\,pc centred on the Sun, and select objects via a random forest classification algorithm using various parameters defined in \textit{Gaia} such as the renormalised unit weight error and the parallax error. The selected sample is roughly uniform over the sky, as expected, and most rejected objects are in regions of high surface density, also as expected. To extract the white dwarf only sample, another white dwarf random forest classifier was created using the three \textit{Gaia} photometric bands, proper motions, parallaxes and parallax uncertainties. The full catalogue contains over 330\,000 stars, and over 15\,000 white dwarfs within 100\,pc of the Sun.


In line with our other samples, we applied slightly stricter cuts to the GCNS catalogue. We only retained objects with $P_\mathrm{WD-GCNS}\geq0.75$, a distance $\leq$100\,pc, and a white dwarf mass between $0.54-1.34$\,M$_{\odot}$ to eliminate double degenerate candidates and physically implausible single-star evolved white dwarfs. This left us with 14\,699 white dwarf candidates.


\subsubsection{GSPC-WD sample}\label{sec:Vincent}

A new feature of \textit{Gaia} DR3 compared to DR2 or EDR3, is the inclusion of flux-calibrated low resolution spectrophotometry for over 200 million sources \citep{GaiaDR3_2023}. The raw spectra in the blue (BP) and red (RP) photometric channels are calibrated, then merged into a single XP spectrum \citep{Montegriffo_2023, DeAngeli_2023}. \cite{Vincent_2024} used an apparent magnitude limited (\textit{G}$<20.5$) catalogue of white dwarfs with XP spectra (the \textit{Gaia} Synthetic Photometry Catalogue for White Dwarfs, GSPC-WD). A machine learning based classification system gave each spectrum a spectral type and composition, from which white dwarf parameters could be calculated using SDSS-like photometry generated from \textit{Gaia} XP spectra, astrometry and tailored model atmospheres. The catalogue's completeness follows that of \cite{GentileFusillo_2021} up to about 50\,pc, beyond which the completeness of the GSPC-WD drops below that of Gentile Fusillo's catalogue, with an effective magnitude limit of about $G \approx20$. There is also evidence of a colour-dependent decrease in completeness towards redder colours in the GSPC-WD \citep{JimenezEsteban_2023}. By using this catalogue, the number of white dwarfs with known spectral type increases by a factor of about 3 within 100\,pc \citep{Vincent_2024}.

By keeping only objects with a probability of being a white dwarf greater than 0.75, limiting the sample to only those within 100\,pc and limiting objects to those with a white dwarf mass between $0.54-1.34$\,M$_{\odot}$ to reduce the impact of binary white dwarfs, the GSPC-WD contains 13\,235 white dwarfs with spectral classifications, masses, $T_{\rm eff}$, $\log g$, and more, which we use directly from the \citet{Vincent_2024} source catalogue. We maintain the white dwarf probability cut despite these objects having white dwarf spectral types to keep the selection criteria the same across all samples. This removes less than 1 per cent of their catalogue.

Because DC white dwarfs below 5000\,K have unconstrained compositions, we assume a hydrogen composition for these objects in the sample. While the white dwarf sample is smaller than the previous two defined so far, it is the only all-sky sample discussed in this work for which the masses and effective temperatures are based on the appropriate atmospheric composition, while the previous two samples assume pure-hydrogen atmospheres throughout.

\subsection{Part-Sky 100\,pc Samples}

\begin{figure*}
	\includegraphics[width=2\columnwidth]{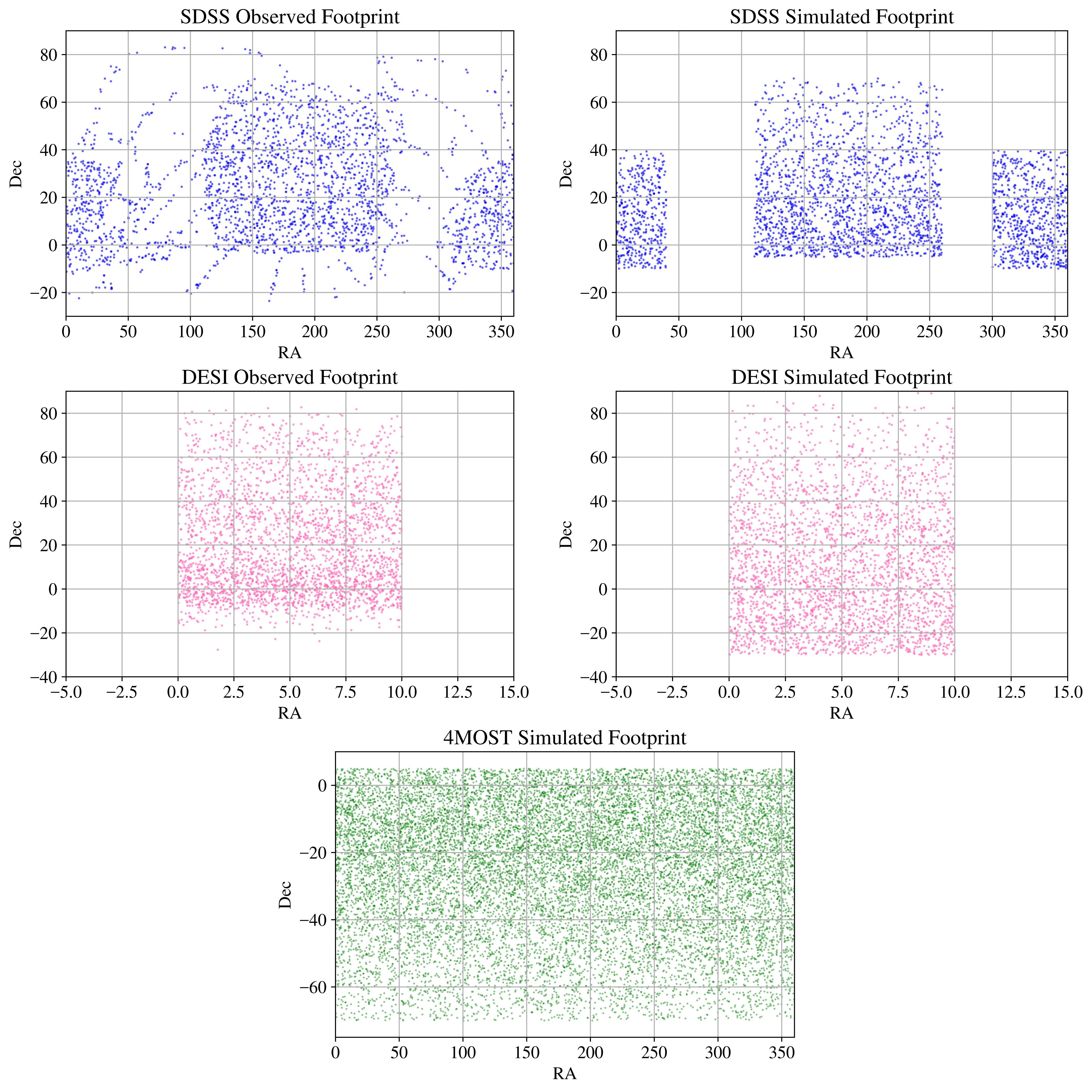}
    \caption{\textit{Top:} The footprint in RA and Dec (degrees) of the SDSS sample taken from \protect\cite{Kilic_2025_100pc}, and the simplified footprint used to generate our simulation of the SDSS sample. Our intention was not to match the footprints completely, but mostly simulate the same parts of the sky. \textit{Middle:} The footprint in RA and Dec of the DESI sample taken from \protect\cite{DESI_2025}. 
    \textit{Bottom:} The simulated 4MOST footprint based on the planned sky coverage of the instrument.}
    \label{fig:footprints}
\end{figure*}

Not all surveys cover and observe the whole of the sky; for instance ground-based MOS surveys are restricted based on the telescope location, and often have limited footprints, for instance by focusing (or avoiding) the Galactic plane regions as driven by other Galactic (or extra-galactic) science cases. 
Furthermore, ground MOS surveys typically have a lower completeness than photometric surveys, owing to both the instrumental limitations in crowded fields, and the pressure factor of different science sub-surveys trying to obtain limited fibre time. 

By comparing these part-sky 100\,pc samples to the previously defined all-sky samples, we assess whether the MOS datasets can be used in the same way to derive white dwarf statistical properties and star formation histories.

Specifically, we examine three white dwarf spectroscopic samples: that of \cite{Kilic_2025_100pc} based on the SDSS I-IV footprint, the first DESI data release \citep{DESI_2025}, and the predicted white dwarf sample expected to be available after the completion of the 4MOST 100PC and white dwarf sub-surveys around 2031 \citep{4MOST}.

\subsubsection{SDSS}
\label{sec:SDSS}

\cite{Kilic_2025_100pc} use SDSS DR9 and Pan-STARRS DR1 photometry matched with \textit{Gaia} DR3 to create a catalogue of white dwarfs within 100\,pc. They focussed on reliability rather than completeness in their sample, and on spectroscopic follow up. By combining previous literature spectral classifications and follow up observations conducted in their work, the sample has 75 per cent spectroscopic completeness for white dwarfs identified in the SDSS DR9 footprint. Their completeness reaches 86 per cent for $T_{\rm eff} > 5000$\,K, illustrating a bias against cooler, redder white dwarfs similarly to the GSPC-WD sample.

Of the 3145 objects in the sample, we keep 2750 once we have selected only objects with a probability of being a white dwarf greater than 0.75, and only those with a white dwarf mass between $0.54-1.34$\,M$_{\odot}$. The probability cut is kept despite objects in this sample being confirmed white dwarfs to keep the selection criteria the same for all samples used. The $P_{\rm WD}$ used for each object is that of \citet{GentileFusillo_2021}. The sky footprint of this sample can be seen in the top left of Fig.\,\ref{fig:footprints} compared with the approximated analytical footprint used in our population synthesis simulations in the top right. 

\subsubsection{DESI}

The Dark Energy Spectroscopic Instrument (DESI) Data Release~1 \citep{DESI_2025} provides medium-resolution spectra ($R \approx$ 2000--5000) for around 18\,million unique objects and is mainly planned to observe distant galaxies and quasars to explore the nature of dark energy. However, during periods affected by lunar light contamination or twilight, the survey will target nearby bright galaxies and stars, including white dwarfs, as part of its `bright' programme \citep{Cooper_2023}. Using an analogue to the \textit{Gaia} DR2 catalogue of \citet{GentileFusillo_2019} as input catalogue, DESI chooses white dwarf targets largely at random, within its field of view and magnitude limits, immediately prior to each observation. This results in a less biased spectroscopic target selection \citep{Manser_2024} than the SDSS I–IV catalogues.

To define the DESI DR1 100\,pc white dwarf sample, we downloaded the full DESI DR1 dataset\footnote{\url{https://data.desi.lbl.gov/public/dr1/}} and performed a cross-match with the \textit{Gaia} EDR3 catalogue of \citet{GentileFusillo_2021} using the unique \textit{Gaia} DR2 identifier. This yielded a total of 43\,440 unique sources with at least one DESI spectrum. Keeping only objects within 100\,pc, with a probability $P_{\rm WD}\geq 0.75$, and within the mass range $0.54-1.34$\,M$_{\odot}$, we obtained a final sample of 2978 white dwarfs. The sky footprint of this sample can be seen in the middle left of Fig.\,\ref{fig:footprints}, compared with our simulated footprint in the middle right.

\subsubsection{4MOST}

The 4-metre Multi-Object Spectroscopic Telescope (4MOST) is a project by the European Southern Observatory to provide the VISTA telescope with a fibre-fed spectroscopic survey facility, with first light in October 2025. The aim is to provide complementary data for large scale sky surveys such as \textit{Gaia}, Euclid, and PLATO. 
With the knowledge that 4MOST plans to survey most of the southern sky, we can begin to predict the data to be obtained. 4MOST will survey all RA values, but will be limited to between 5 to $-$80 degrees in Dec. The expected and simulated footprint of the 4MOST sample can be seen at the bottom of Fig.\,\ref{fig:footprints}.

The 4MOST 100PC sub-survey (PI Tremblay), within the S3 - Milky Way Bulge and Disk Low Resolution Survey (4MIDABLE-LR) survey \citep{4MIDABLE}, is dedicated to obtaining optical medium-resolution ($R \approx$ 5000) spectra of $\approx$200\,000 \textit{Gaia}-selected main-sequence stars and $\approx$8000 white dwarfs within the 4MOST footprint and 100\,pc of the Sun. One goal of the survey is to compare the SFH derived from stars and white dwarfs, and calibrate stellar ages. There is also a 4MOST white dwarf sub-survey (PIs Tremblay/Gentile Fusillo) set to observe white dwarfs at all distances. The aim is for a signal-to-noise (S/N) of 30 at 4500\,\AA\ for $G < 18.5$\,mag, and a S/N of 15 at $18.5 < G < 19.5$. 4MOST will observe even fainter targets from its ancillary catalogue, but we use an effective $G=19.5$\,mag cut-off corresponding to high-quality (high S/N) data.

\section{Methods}\label{sec:methods}

\subsection{Simulation}\label{sec:sim}

We use the single-star population synthesis code of \citet{Roberts_2025} to generate populations of stars in the local region of the Galaxy according to the initial mass function, a given SFH, and a spatial distribution within the disc. They are then evolved kinematically and temporally, where it then selects the population of white dwarfs within 100\,pc at the present day. The code generates the luminosities, magnitudes, and other observational properties of all the white dwarfs in the simulation to compare against the observed distributions of the samples discussed in the previous section. 
The simulation code takes the same four assumed approximate SFH forms as \citet{Roberts_2025}, shown in Fig.\,7 of that paper. Each ingredient in the simulation introduces an uncertainty into the final predicted ages, masses, luminosities, or magnitudes of the white dwarfs. 

The simulation uses: a population age of $10.6\pm0.5$\,Gyr \citep{Cukanovaite_2023}; an initial mass function index of $2.35\pm0.075$ \citep{Salpeter_1955}; an initial mass range of $0.94-6.84\,\rm M_{\odot}$ with no uncertainty as it was derived self consistently from the 40\,pc sample \citep{Cunningham_2024}; an initial-final mass relation with no uncertainty as it was derived self consistently from the 40\,pc sample \citep{Cunningham_2024}; an initial metallicity of $\rm Z = \rm Z_{\odot} = 0.0134\pm0.0104$ \citep{Asplund_2009}; main sequence and giant phase lifetimes from \cite{Byrne_2024} with a multiplicative uncertainty of $1\pm0.035$; merger delays from \cite{Temmink_2020}; a He-atmosphere white dwarf fraction of $0.25\pm0.065$ \citep{OBrien_2024}; an age-kinematics relation using a scale height of $10.71\times\rm age_{ total}[Gyr] + 65$ from \cite{Cukanovaite_2023} with a multiplicative uncertainty on the slope of $1\pm0.3$; cooling models from \cite{Bedard_2020} with a multiplicative uncertainty of $1\pm0.045$; and a crystallisation cooling delay of $0.5$\,Gyr \citep{Blouin_2019, Kilic_2020}. We refer the reader to \citet{Roberts_2025} Tables\,1 and 3 for further details on these ingredients and uncertainties.


Some of these uncertainties are systematic (population age, initial mass function, main sequence lifetimes, spectral evolution, merger delays, age vs kinematic relation, cooling ages, crystallisation delays) and some affect individual stars in each simulation (metallicity, and any \textit{Gaia} data errors associated with each star). 

This work focusses on two of the three methods used in \citet{Roberts_2025}, comparisons of the white dwarf luminosity function and absolute \textit{Gaia G} magnitude distribution, with the third - the direct age method -  explored only in Appendix A for those who wish to see all three methods tested on the 100\,pc samples. The error bars on all simulations are generated by running the simulations 100 times allowing all ingredients to vary within their uncertainties and taking the mean and standard deviation of the values in each luminosity or absolute magnitude bin.

For part-sky samples, we only keep the white dwarfs found in the relevant footprints described by Fig.\,\ref{fig:footprints}. 
Restricting to survey footprints can amplify the influence of Galactic structure effects, namely the Galactic scale height and the age-velocity relation. A part-sky survey pointing towards the Galactic centre will not see the Galactic scale height, but a part-sky survey pointing out of the Galactic plane will be far more susceptible to the scale height, the age-velocity relation, and the assumptions underlying their modelling. 

\subsection{Luminosity function}\label{sec:LF}

To calculate the bolometric magnitudes of white dwarfs in the sample, the effective temperature and radius of the white dwarf are required. This itself requires knowing the atmospheric composition of the white dwarf, i.e. hydrogen- or helium-dominated, as fitting temperature and radius requires fitting a model atmosphere and mass-radius relation to the photometry and astrometry (or, although not used in this case, fitting the same models to spectroscopy). With the 40\,pc sample, all of this composition information is known, but for the 100\,pc samples used in this work, this is only known for the samples of \citet{Vincent_2024} and \cite{Kilic_2025_100pc}. Because of this, we chose to use temperatures and masses as derived from pure-hydrogen atmospheres for other samples lacking spectroscopic classifications or with incomplete spectroscopic coverage. 
Fig.\,\ref{fig:100vs40pc_results} demonstrates that this assumption makes little difference, as the derived 100\,pc white dwarf luminosity functions are very similar to the one for the volume complete 40\,pc sample, which has full atmospheric composition information. The relative contribution of stars that would be affected by this assumption is as follow: in \cite{Vincent_2024}, where full spectroscopic information is available for helium-atmosphere DB/DBA, warm DC (> 5000 K) and DZ/DQ types, the percentage of helium white dwarfs that could have significantly affected masses and temperatures is 10.1\%

We note that most cool white dwarfs ($T_{\rm eff}<5000\,$K) are of DC spectral type with unconstrained compositions \citep[e.g.][]{Elms_2022}, hence the assumption of pure-hydrogen atmospheres is actually done for all samples at the faint end of the luminosity function, regardless of the availability of spectroscopy.


On the other hand, in the simulated luminosity functions, all samples without composition information rely on the 25 per cent helium fraction as described in Sect.\,\ref{sec:sim} and each white dwarf is randomly assigned a composition. For the samples of \citet{Vincent_2024} and \citet{Kilic_2025_100pc}, we can predict a slightly more accurate luminosity function using the known compositions of each white dwarf. We find that these two options make very little difference in the predicted luminosity functions.

\begin{figure}
	\includegraphics[width=0.9\columnwidth]{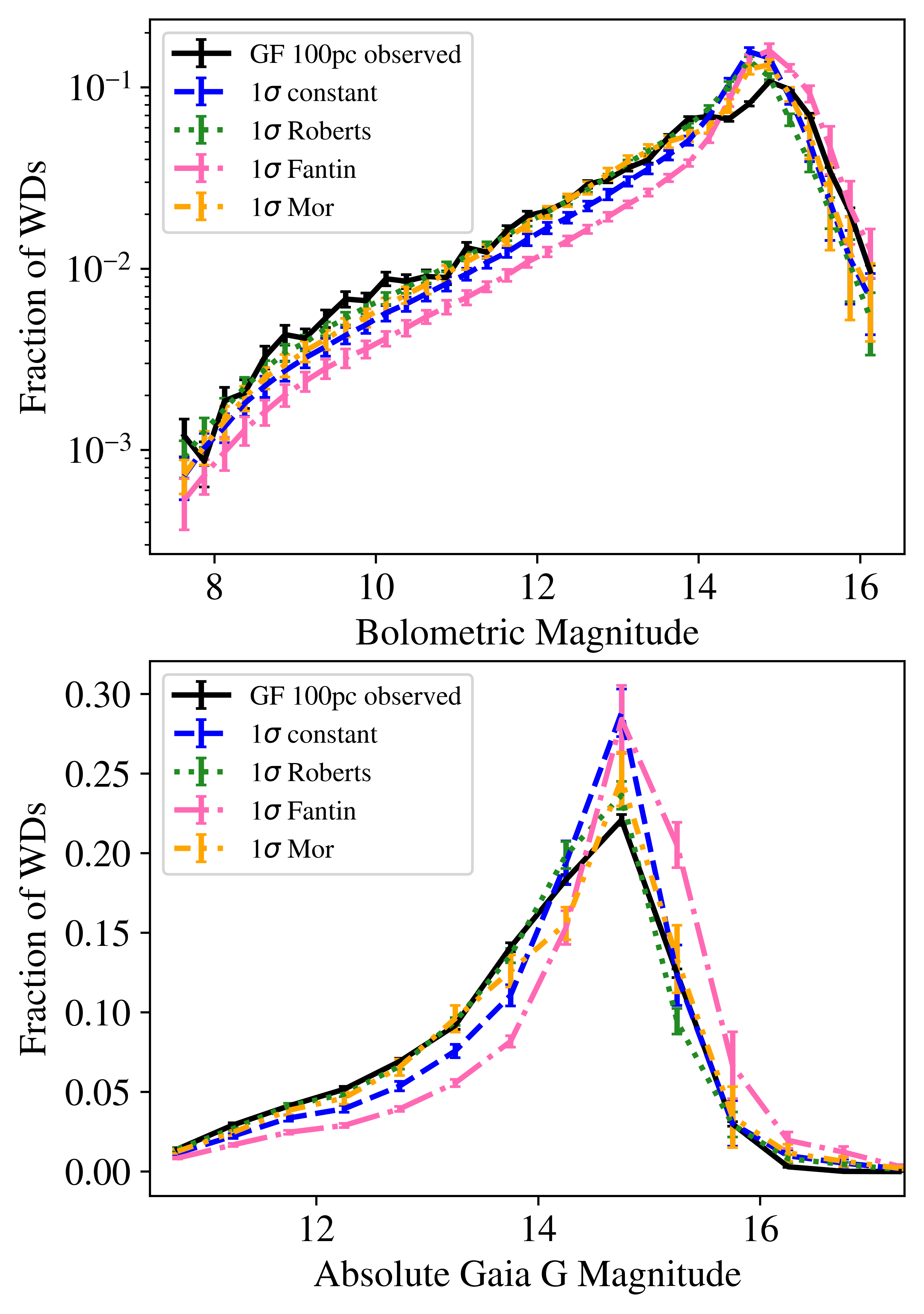}
    \caption{\textit{Top:} Results from the luminosity function method with population synthesis using all four assumed star formation forms (constant; \citealt{Roberts_2025} direct age method; \citealt{Fantin_2019}; \citealt{Mor_2019}) compared to observed 100\,pc white dwarf sample of \citet{GentileFusillo_2021}. Error bars are the standard deviation of 100 runs of the simulations, and Poisson errors for the observed sample.  \textit{Bottom:} Results from the absolute \textit{Gaia G} magnitude method with all four assumed star formation forms compared to the observed 100\,pc sample of \citet{GentileFusillo_2021}. Error bars follow the same convention as above.}
    \label{fig:GF_results}
\end{figure}

\subsection{Absolute \textit{Gaia G} magnitude distribution}\label{sec:absG}

Deriving the observed absolute \textit{Gaia G} magnitude distribution only requires a parallax and an apparent \textit{Gaia G} magnitude, both of which are by definition available for all our \textit{Gaia}-sourced catalogues. Using this technique, all information about spectral types, atmospheric compositions, hence cooling times and atmospheric parameters, is encoded in the simulated population. The simulation relies on composition statistics from the smaller 40\,pc sample, assuming the breakdown is the same for a larger volume, and assumes that 25 per cent of the white dwarf population has a helium-rich atmospheric composition, assigned randomly. 


All observed magnitudes are dereddened based on either the $V$-band extinction ($A_V$) for that object in \citet{GentileFusillo_2021}, or if missing, an $A_V$ extinction value sampled from a Gaussian distribution with the mean as 0.02\,mag (and standard deviation of 0.015\,mag), which is the average extinction over the whole 100\,pc sample \citep{GentileFusillo_2021,Sahu_2024}.

\section{Results}\label{sec:results}


\subsection{Gentile Fusillo et al sample}

Figure\,\ref{fig:GF_results} shows the comparison of the observed $G_{\rm abs}$ and luminosity functions with the simulated samples of white dwarfs. For the luminosity function simulations, the best fitting SFH is that of \cite{Roberts_2025} based on the 40\,pc sample of white dwarfs, followed by \cite{Mor_2019}. The constant SFH \citep{Cukanovaite_2023} and that of \cite{Fantin_2019} struggle to reproduce the observations, in particular around the rising slope and peak of the function. It should be noted that none of the assumed SFH forms fit particularly well around the peak, with all of them predicting a small upturn just below $M_{\mathrm{bol}}=15$ that is not observed. 

The population age for all four simulations was allowed to vary as a Gaussian around 10.6\,Gyr with a standard deviation of 0.5\,Gyr and this gives a good agreement with the downturn of the luminosity function of Gentile Fusillo's sample regardless of the SFH chosen. The age that gives the best result is around 11.0\,Gyr but this is within $1\sigma$ of the default value. This suggests that the 100\,pc \textit{Gaia} sample is reasonably complete and unbiased compared to the 40\,pc sample, as it agrees within 1$\sigma$ with the same age and SFH.

The absolute \textit{Gaia G} magnitude distribution shows many of the same trends, namely an over-prediction of stars at the peak which is not seen in the observations. Again, both \cite{Roberts_2025} and \cite{Mor_2019} show the best match, while a constant SFH and that of \cite{Fantin_2019} are noticeably poorer. The early-peaking SFH of Fantin in particular produces too many faint stars and too few bright stars when compared to the observed 100\,pc sample, even accounting for uncertainties. 


\subsection{Gaia Catalogue of Nearby Stars sample}

\begin{figure}
	\includegraphics[width=0.9\columnwidth]{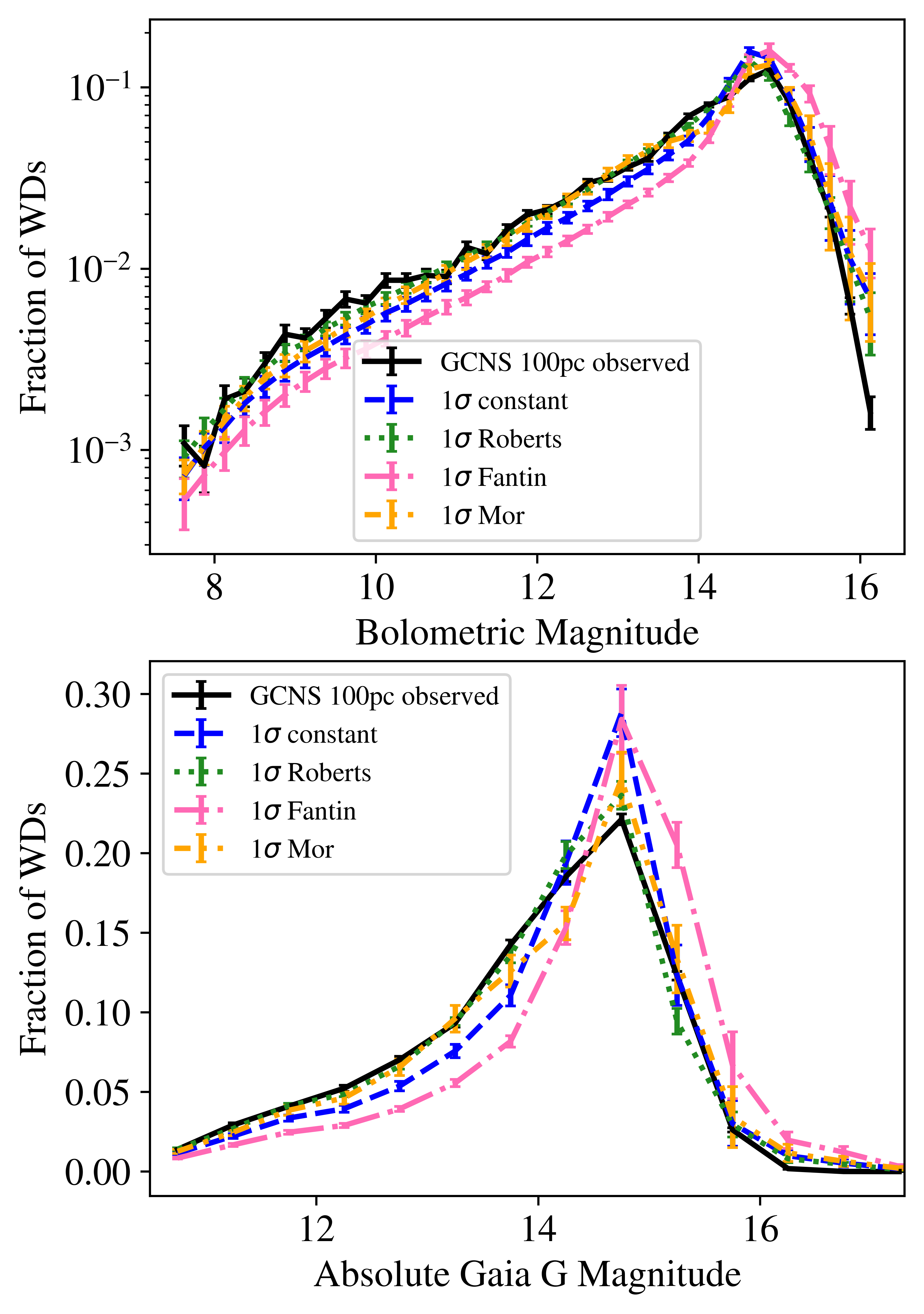}
    \caption{Similar to Fig.\,\ref{fig:GF_results}, but for the 100\,pc GCNS white dwarf catalogue \citep{GCNS_2021}.}
    \label{fig:GCNS_results}
\end{figure}

Due to the similarities between the GCNS and the catalogue of Gentile Fusillo (see Fig.\,\ref{fig:100vs40pc_results}), drawn from the same underlying \textit{Gaia} DR3 catalogue, it is no surprise in Fig.\,\ref{fig:GCNS_results} that the GCNS sample also has a best match to the star formation histories of \cite{Roberts_2025} followed by \cite{Mor_2019} and \cite{Cukanovaite_2023}. Once again, the GCNS does not show the upturn at the peak that the simulation suggests should appear.

The population age for all four simulations was allowed to vary as a Gaussian around 10.6\,Gyr with a standard deviation of 0.5\,Gyr and this gives a good agreement with the downturn of the luminosity function of the sample across all four star formation histories, justifying our choice of this value. In fact, the best fitting age for this sample is 10.6\,Gyr.



\subsection{GSPC-WD sample}

\begin{figure}
	\includegraphics[width=0.9\columnwidth]{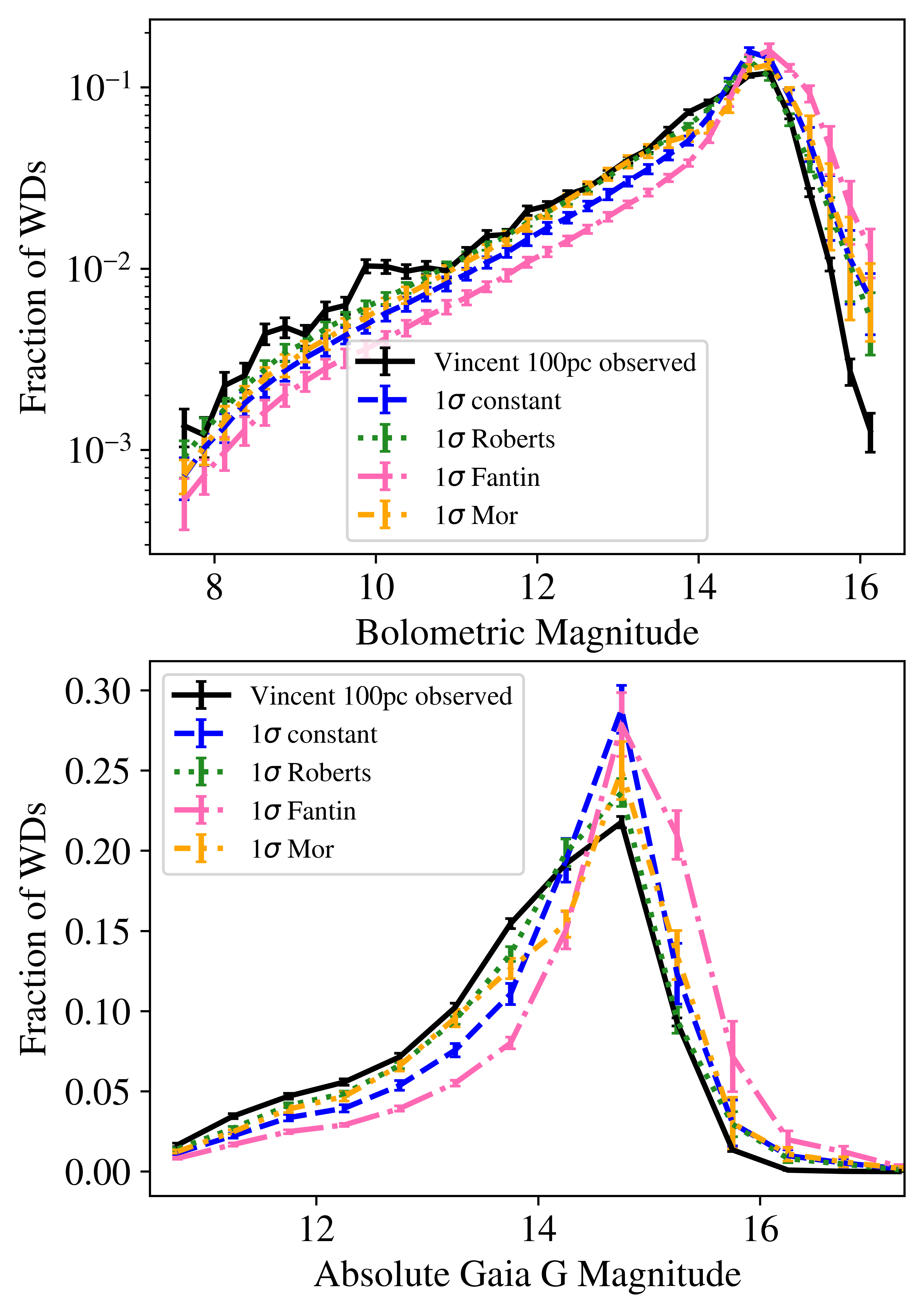}
    \caption{Similar to Fig.\,\ref{fig:GF_results}, but for the 100\,pc GSPC-WD catalogue \citep{Vincent_2024}.}
    \label{fig:Vincent_results}
\end{figure}

Figure\,\ref{fig:Vincent_results} shows that the GSPC-WD sample best fits the SFH of \cite{Roberts_2025}, then \cite{Mor_2019}, \cite{Cukanovaite_2023}, and finally \cite{Fantin_2019}. 
However the luminosity function does reveal the extent of the incompleteness of the sample, as described in Sect\,\ref{sec:Vincent} and by the authors themselves. The downturn of the luminosity function constrains the age of the sample
and in order to match the downturn of the Vincent sample, the population age would need to change from a default 10.6\,Gyr to a maximum of 9.3\,Gyr for the constant SFH, all the way down to 8.5\,Gyr for the \cite{Fantin_2019} formation history. This is just within the $3\sigma$ uncertainty of our population age \citep{Roberts_2025}, and shows that the incompleteness of the GSPC-WD sample does indeed significantly bias against fainter and redder white dwarfs. 

\subsection{SDSS}

The sample from \cite{Kilic_2025_100pc} behaves similarly to the previous 100\,pc samples in that the formation histories from \cite{Roberts_2025} and \cite{Mor_2019} fit the best.
As with the GSPC-WD sample, the luminosity function does reveal the extent of the incompleteness of the sample, as described in Sect\,\ref{sec:SDSS} and by the authors.
The assumed age of the population needs to be reduced from 10.6\,Gyr to anywhere between 9.3 to 7.5\,Gyr, depending on the selected SFH, making it increasingly less helpful for constraining the underlying SFH of the 100\,pc volume. Additionally, the bias of the sample is not necessarily just against very faint white dwarfs; the gradient of the drop off in both the luminosity function and the absolute \textit{Gaia G} magnitude distribution does not match the gradient in any simulation, 
possibly due to the authors avoiding the spectroscopic follow-up of objects more likely to be featureless DC white dwarfs below 5000\,K.



\begin{figure}
    \includegraphics[width=0.9\columnwidth]{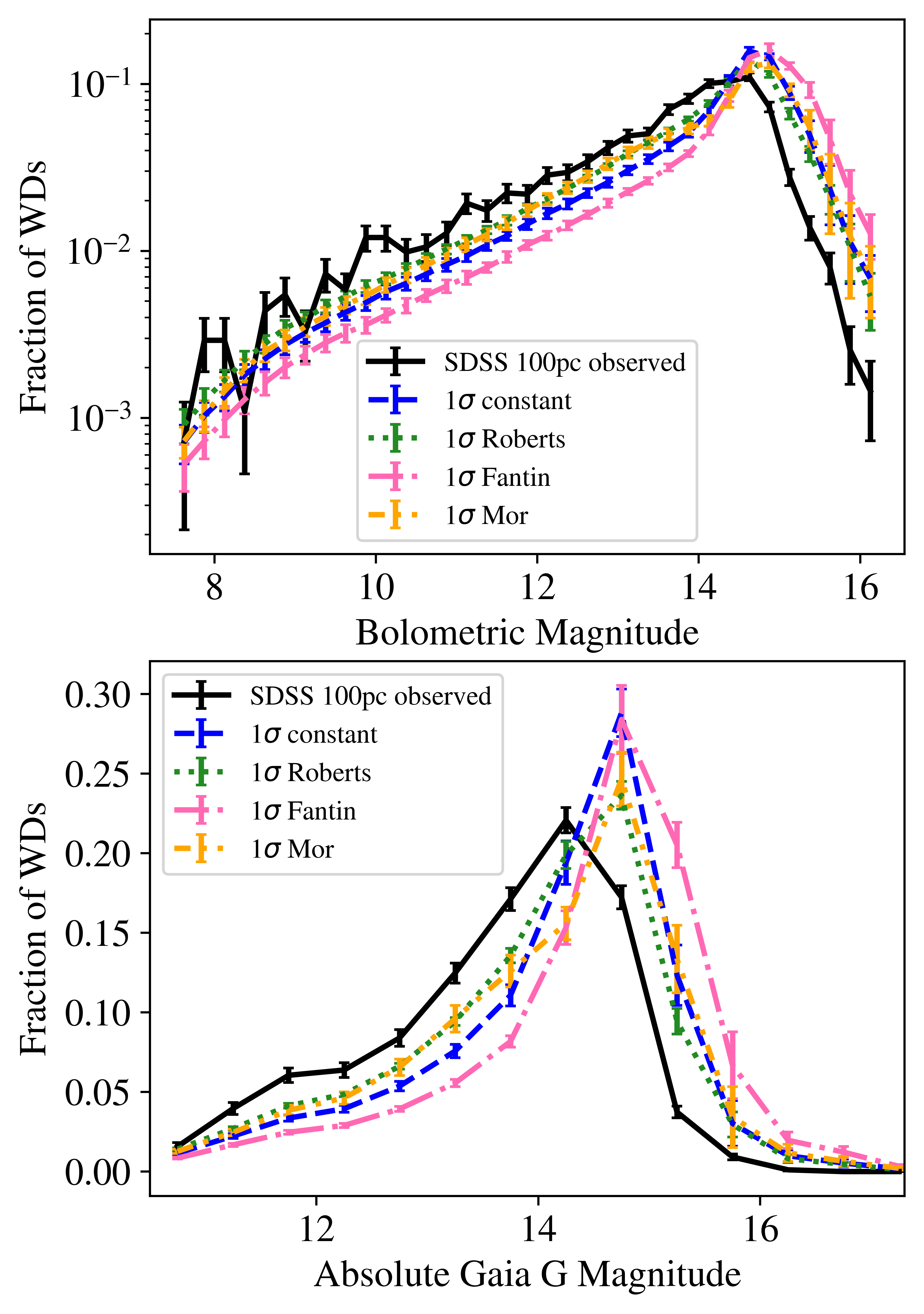}
    \caption{Similar to Fig.\,\ref{fig:GF_results}, but for the part-sky 100\,pc SDSS white dwarf catalogue of \citep{Kilic_2025_100pc}.}
    \label{fig:SDSS_results}
\end{figure}

\subsection{DESI}

\begin{figure}
    \includegraphics[width=0.9\columnwidth]{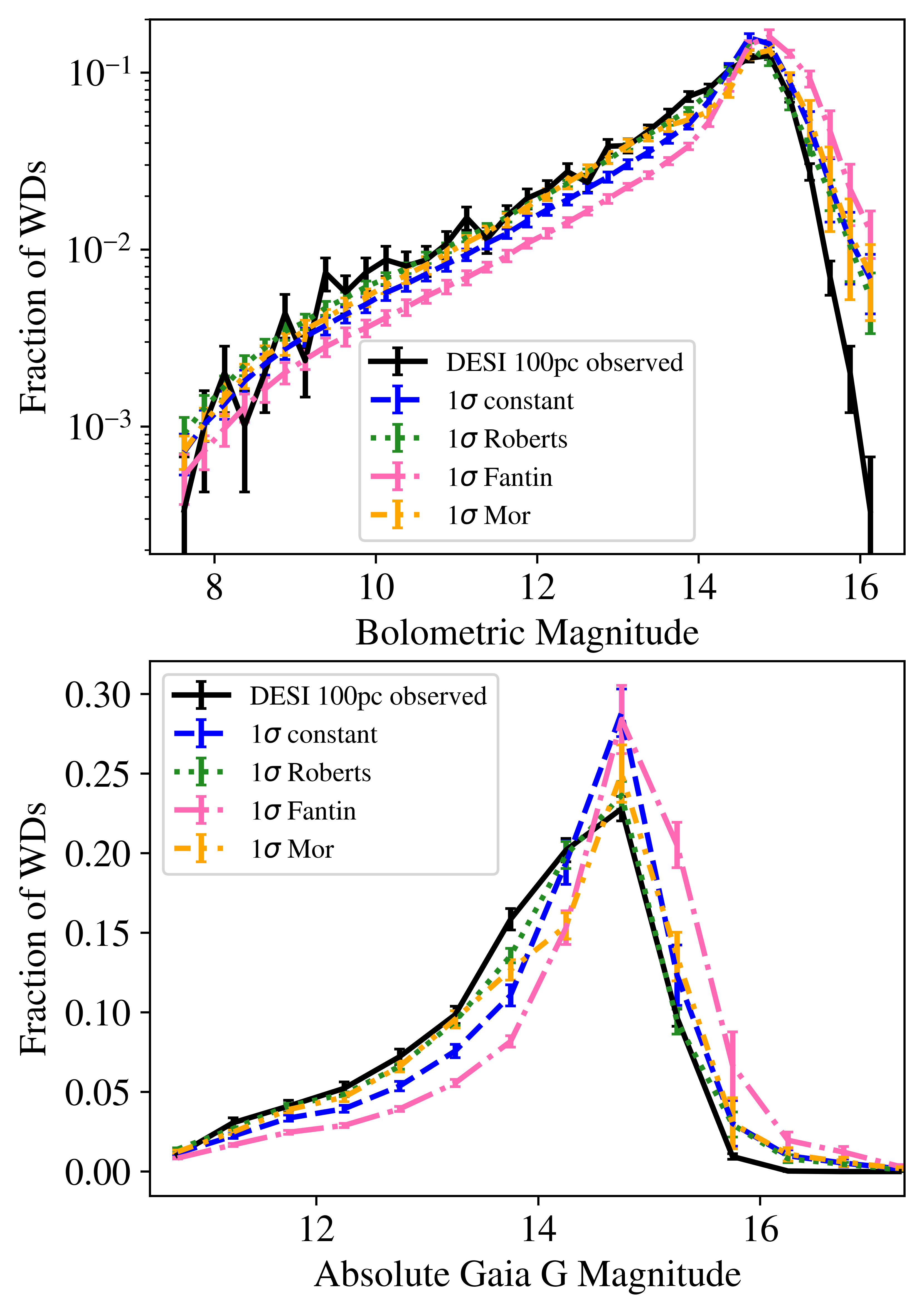}
    \caption{Similar to Fig.\,\ref{fig:GF_results}, but for the 100\,pc DESI white dwarf catalogue \citep{DESI_2025}.}
    \label{fig:DESI_results}
\end{figure}

The part-sky DESI DR1 sample agrees best with the SFH we found in \citet{Roberts_2025} from the 40\,pc sample as well as \citet{Mor_2019}.
The DESI sample does not suffer from the same apparent incompleteness as SDSS and GSPC-WD in either distribution: the downturn of the luminosity function is well fitted by all four star formation histories using a population age of 10.6\,Gyr.

We theorise that this seemingly better completeness is in part due to the random nature of DESI's sampling; a location is picked at random within the footprint, with uniform distribution of tiles over the sky \citep{Schlafly_2023}, and, once the predefined targeting criteria are applied, the target selected there is also random, so there is no bias to brighter or fainter objects within the white dwarf sample. The DESI white dwarf survey targets objects mostly from \citet{GentileFusillo_2019}'s sample \citep{Cooper_2023}, which is established to be largely volume complete and unbiased within 100\,pc, as discussed in previous sections. Selection is made with photometric measurements from \textit{Gaia} DR2 and astrometric measurements from \textit{Gaia} EDR3 and is based on equations (1)-(7) in \citet{GentileFusillo_2019}. 99 per cent of the DESI white dwarf candidates are expected to be given a fibre during the main survey but the choice of which white dwarfs will be selected is decided right before an observation is made and therefore is not predictable \citep{
Manser_2024}. Overall, this gives all white dwarfs an equal chance of being observed, and as such gives the sample low bias.

\begin{figure}
    \includegraphics[width=0.9\columnwidth]{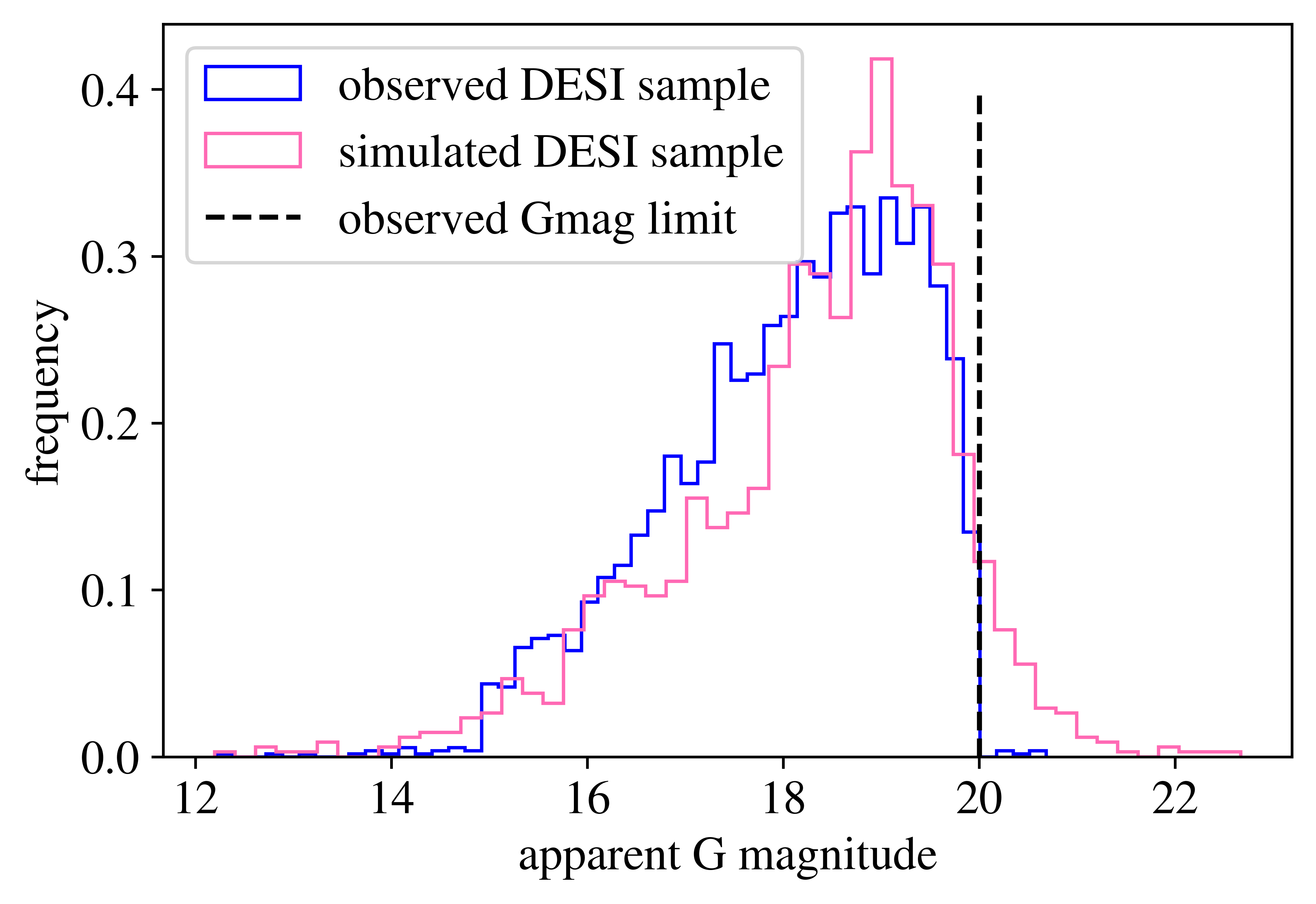}
    \caption{Observed apparent $G$ magnitude distribution of the DESI 100\,pc sample and the 100\,pc population simulation for the DESI footprint. The empirical apparent magnitude limit of $G=20$ (dashed line) does exclude about 5 percent of the simulated population.}
    \label{fig:mag_lim}
\end{figure}

It appears that the DESI DR1 sample and future data releases could be used as analogous to all-sky, volume-complete samples. If we simulate DESI as a full sky sample and compare against the observed part sky sample, it remains a good fit. In an all-sky sample, we predict that DESI would see more very faint stars, decreasing the gradient of the drop off at $M_\mathrm{bol} \approx 15$. Overall though, DESI's observational footprint matches the age and SFH of full sky samples. 
The DESI DR1 white dwarf sample has an empirical apparent magnitude limit of $G = 20$, and the population synthesis code does not currently make any cuts at magnitude limits since it recreates volume limited samples. The empirical apparent magnitude limit of DESI and its comparison to the simulation can be seen in Fig.\,\ref{fig:mag_lim}. About 5 per cent of the simulated sample is below that magnitude limit, which makes the observed DESI luminosity function steeper than the simulations.

\subsection{4MOST}

Without an observational sample to compare against, we aim to provide a simple prediction of what to expect from the 4MOST 100PC sub-survey by the end of the survey around 2031. 
Our simulations account for the part-sky coverage and the 4MOST 100PC sub-survey magnitude limit of $G \leq 19.5$, but neglect the fibre-fed spectroscopic selection function. As with DESI, 4MOST will select white dwarfs randomly from the input catalogue of \citet{GentileFusillo_2021}, so the bias from the selection function is expected to be small. The key difference is that 4MOST relies on the more complete \textit{Gaia} DR3, whereas DESI used DR2. Not all white dwarfs within the footprint are expected to be selected, primarily because the significantly smaller fibre spatial resolving power compared to \textit{Gaia} excludes targets near neighbouring sources, but also because of time-pressure from other sub-surveys. However, since the white dwarf selection operates in a largely random fashion, we only expect an overall reduction in completeness relative to a truly volume-limited sample (by a few percent up to $\approx$30 per cent), with no current evidence for colour-, magnitude-, or distance-dependent biases, except for the previously mentioned apparent magnitude limit.

\begin{figure}
    \includegraphics[width=0.9\columnwidth]{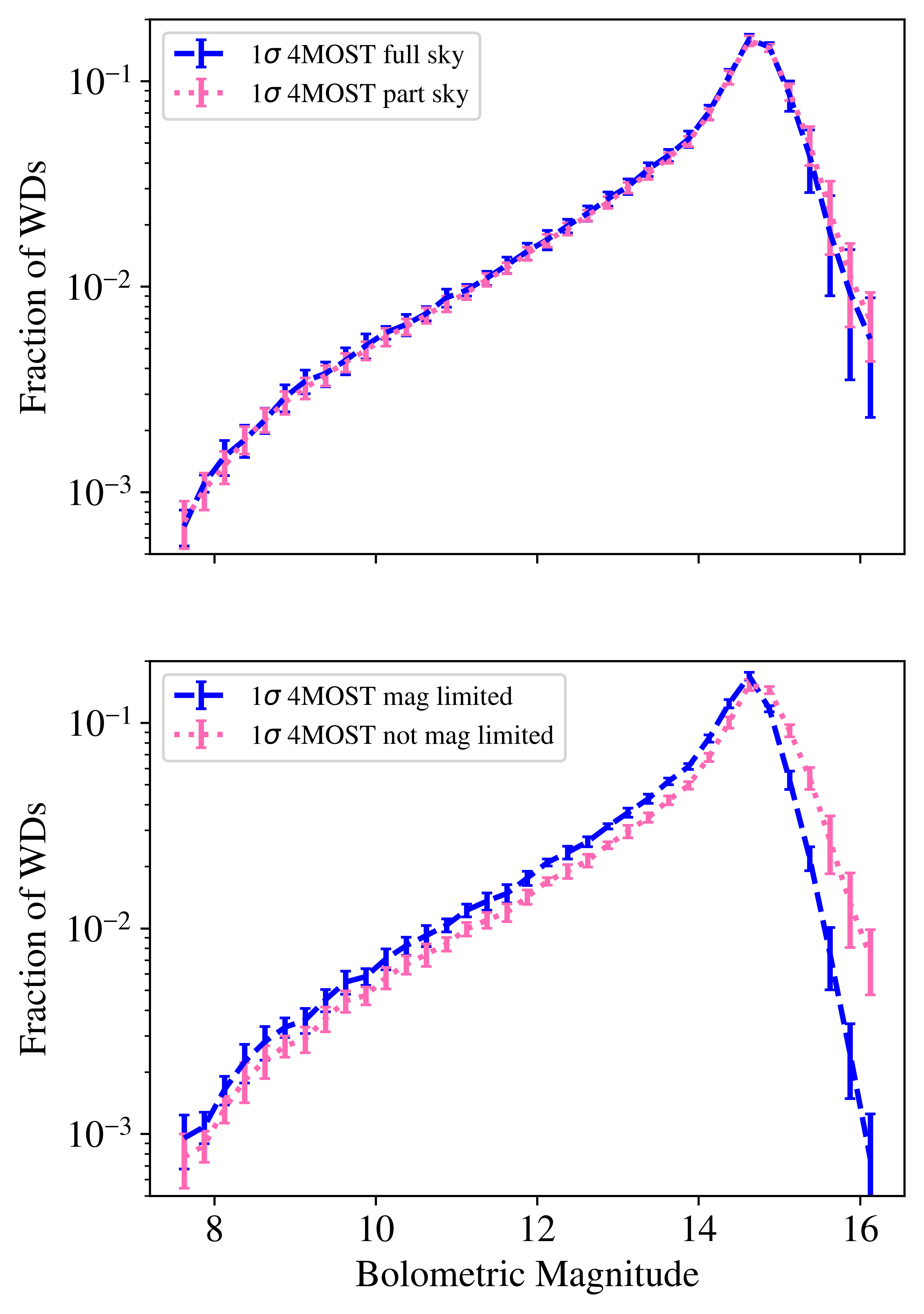}
    \caption{\textit{Top:} The predicted luminosity function of the 4MOST 100PC sub-survey using its planned part-sky footprint as shown in the bottom panel of Fig.\,\ref{fig:footprints}. This is compared against a full sky sample with the same magnitude limit. There is no predicted change in the luminosity function due to 4MOST being a part-sky survey.
    \textit{Bottom:} The predicted luminosity function of the 4MOST 100PC sub-survey using its planned part-sky footprint, with and without its planned 19.5 apparent magnitude limit. The magnitude limit does cause some incompleteness at the faint end of the luminosity function.}
    \label{fig:completeness}
\end{figure}

\begin{figure}
    \includegraphics[width=0.9\columnwidth]{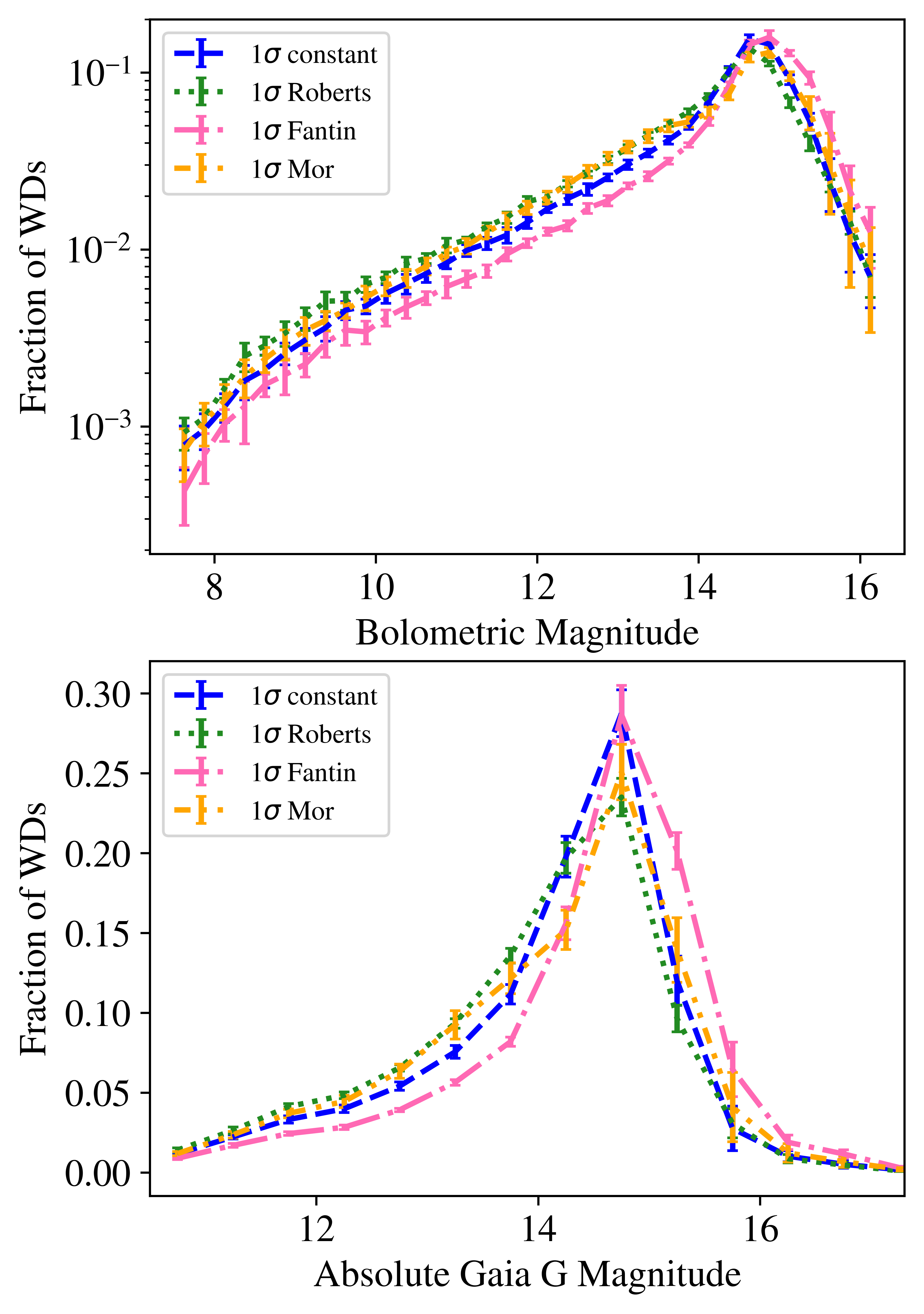}
    \caption{Similar to Fig.\,\ref{fig:GF_results}, but for the predicted 100\,pc 4MOST white dwarf catalogue \citep{4MOST}.}
    \label{fig:4MOST_results}
\end{figure}

Figure\,\ref{fig:completeness} shows that the part-sky selection makes no significant difference in the luminosity function within errors, which means that 4MOST could be used alongside all-sky samples without the need to account for biases based on the smaller sky area observed. 
This conclusion is bolstered by the similar results found with the DESI DR1 sample in the previous section. The magnitude limit of the 4MOST 100PC sub-survey does cause some incompleteness at the faint end of the luminosity function, as shown in the bottom panel of Fig.\,\ref{fig:completeness}. This suggests that accurately determining the age of the Galactic disc will remain challenging in the MOS survey era. In addition to the model age systematics discussed by \citet{Cukanovaite_2023,Roberts_2025}, the oldest and faintest white dwarfs will be missed even within the 100\,pc sample, meaning that understanding the faint end of the luminosity function will be benefit more from deeper surveys, such as Vera Rubin \citep{Fantin_2020}.
Nevertheless, Fig.\,\ref{fig:4MOST_results} demonstrates that 4MOST will be able to distinguish within error bars between the four forms of star formation used in this work.

\section{Discussion and Conclusion}
\label{sec:conc}

It is clear from the previous sections that the best fitting SFH does not depend on the specific \textit{Gaia} 100\,pc selection or sample of white dwarfs used: across all samples tested, the late peaking SFH of \cite{Roberts_2025} fits best and provides consistency with the 40\,pc sample of white dwarfs. This is reassuring: the smaller 40\,pc sample with full spectroscopic follow-up is representative of the larger volume 100\,pc samples, and potentially even larger volume samples still (although larger volumes will need to factor in stellar populations like halo white dwarfs). It is also reassuring that part-sky samples, such as DESI and 4MOST, or samples with no medium-resolution spectroscopic coverage, such as \textit{Gaia} DR3 white dwarf catalogues or future Vera Rubin catalogues, can constrain the local SFH with high accuracy. Previous results for the SFH and age of the population using part-sky samples or samples lacking spectroscopic information or both such as \citet{Rowell_2013} or \citet{Fantin_2019} therefore also should not be biased by these factors and can be compared more easily. Finally, we note that a key limitation of MOS surveys is their magnitude limit, which excludes the coolest white dwarfs even within 100\,pc and therefore prevents significant improvement in constraining the faint end of the luminosity function, or age of the Galactic disc, beyond what is already achievable within $\approx$40\,pc.

The results also show that the assumed scale height of the Galaxy and age-velocity dispersion relation, which become relatively more important as we get further away from the Galactic disc, have been modelled accurately enough for consistency within 100\,pc. As described in \citet{Roberts_2025}, the age dependent scale height of the Galaxy is modelled as $h=10.71(t_{\rm total}) +65~$ where $t_{\rm total}$ is the total age of a star. The age dependent probability of having left the sample is then $p({\rm left}) = 1 - 65/h$. However, this would need to be confirmed for larger volumes.

We show that a late peaking star formation rate such as \cite{Roberts_2025} or \cite{Mor_2019}, or a constant star formation rate, are the best fits overall to the 100\,pc samples investigated here. While the SFH derived in \citet{Roberts_2025} is the best fitting SFH, the large systematics (detailed in \citet{Roberts_2025}) mean that it is statistically impossible to say which of those three should be favoured. Therefore, we advise that a constant star formation rate can be used as a simple but effective assumption \citep{Cukanovaite_2023}. We also remind the reader that due to the age-velocity dispersion relation, a constant star formation does not result in a uniform stellar age distribution at present day in a volume-complete sample. Furthermore, the age distribution of white dwarfs is further biased by the missing main-sequence stars \citep{Roberts_2025}.

Since the 100\,pc samples investigated here have at least 2.5 as many stars and at most 15 times as many stars as the 40\,pc sample, Poisson noise is a lot smaller and it can be better assumed that the second-order features of the luminosity function and absolute $G$ magnitude distribution are due to underlying star formation histories rather than just random noise. For example in Fig.\,\ref{fig:100vs40pc_results}, a small bump is seen in the 40\,pc absolute \textit{Gaia G} magnitude distribution that is not visible in the full sky 100\,pc samples at a 3$\sigma$ confidence. We can now suggest that this bump is not a feature of the white dwarf population, but noise in the smaller 40\,pc sample. 
However, we note that because of significant model systematics, small features in the observed distributions cannot be reliably attributed to corresponding features (e.g. peaks) in the SFH.

This work only focusses on white dwarfs within the 100\,pc volume and only single white dwarfs within a certain mass range (primarily to exclude binary products and unconfirmed unresolved double degenerate white dwarfs). There is much scope, especially with the 4MOST 100PC sample, to both explore other stellar populations such as the much more numerous main sequence stars, which have also previously been used to investigate the SFH \citep{Mor_2019} and binary white dwarfs. The 100PC 4MOST sample could reveal information on binary mass transfer, mass ratios, ages and many other considerations so far not included in the simulations used here and in \citet{Roberts_2025}. Improving our understanding of these inputs will provide more accurate SFH and age determinations in the future.

\section*{Acknowledgements}

This research received funding from the European Research Council under the European Union’s Horizon 2020 research and innovation programme number 101002408 (MOS100PC). This work has made use of data from the European Space Agency (ESA) mission Gaia (\url{https://www.cosmos.esa.int/gaia}), processed by the Gaia Data Processing and Analysis Consortium (DPAC, \url{https://www.cosmos.esa.int/web/gaia/dpac/consortium)}. Funding for the DPAC has been provided by national institutions, in particular the institutions participating in the Gaia Multilateral Agreement. 
DESI is supported by the Director, Office of Science, Office of High Energy Physics of the U.S. Department of Energy under contract no. DE–AC02–05CH11231, and by the National Energy Research Scientific Computing Center, a DOE Office of Science User Facility under the same contract; additional support for DESI is provided by the U.S. National Science Foundation, Division of Astronomical Sciences under contract no. AST-0950945 to the NSF’s National Optical-Infrared Astronomy Research Laboratory; the Science and Technologies Facilities Council of the United Kingdom; the Gordon and Betty Moore Foundation; the Heising-Simons Foundation; the French Alternative Energies and Atomic Energy Commission (CEA); the National Council of Science and Technology of Mexico (CONACYT); the Ministry of Science and Innovation of Spain (MICINN), and by the DESI Member Institutions: \url{https://www.desi.lbl.gov/collaborating-institutions}. DESI are honoured to be permitted to conduct scientific research on Iolkam Du’ag (Kitt Peak), a mountain with particular significance to the Tohono O’odham Nation.

\section*{Data Availability Statement}

The observational data used in this article are published in \cite{McCleery_2020,GentileFusillo_2021, DESI_2025, OBrien_2024} and \cite{Kilic_2025_100pc}. The results derived in this article will be shared on a reasonable request to the corresponding
author.




\bibliographystyle{mnras}
\bibliography{aamnem99,bib_emily,bpass,bedard} 




\appendix

\section{Direct age calculations}\label{appendix}

The age of each white dwarf in the samples is calculated via:
\begin{equation}
    t_\mathrm{total} = t_\mathrm{WD}+t_\mathrm{MS+GP}+t_\Delta
\end{equation}
{\noindent}where $t_\mathrm{total}$ is the total age of the white dwarf or the time it formed before the present day, $t_\mathrm{WD}$ is the white dwarf cooling time and includes any delays due to crystallisation, $t_\mathrm{MS+GP}$ is the main sequence and giant phases lifetime of the star, and $t_\Delta$ is the length of any evolutionary delays due to the presence of a merger in the star's history (see \citealt{Roberts_2025}). If the inclusion of a merger delay increased the total age of the white dwarf to greater than 10.6\,Gyr, the merger delay was ignored in the calculation of the total age.

As in the technique of the luminosity function, calculating the cooling age requires the effective temperature and radius (or equivalently mass through the mass-radius relation) of the white dwarf, while calculating the main-sequence lifetime requires the white dwarf mass (then processed through the initial-final mass relation). This requires knowing the atmospheric composition of the white dwarf. Assumptions must be made for the \citet{GentileFusillo_2021} and GCNS \citep{GCNS_2021} samples. We assume a hydrogen composition for stars without composition information (see Table~\ref{tab:numbers_in_100pc}). 

\begin{figure}
	\includegraphics[width=0.9\columnwidth]{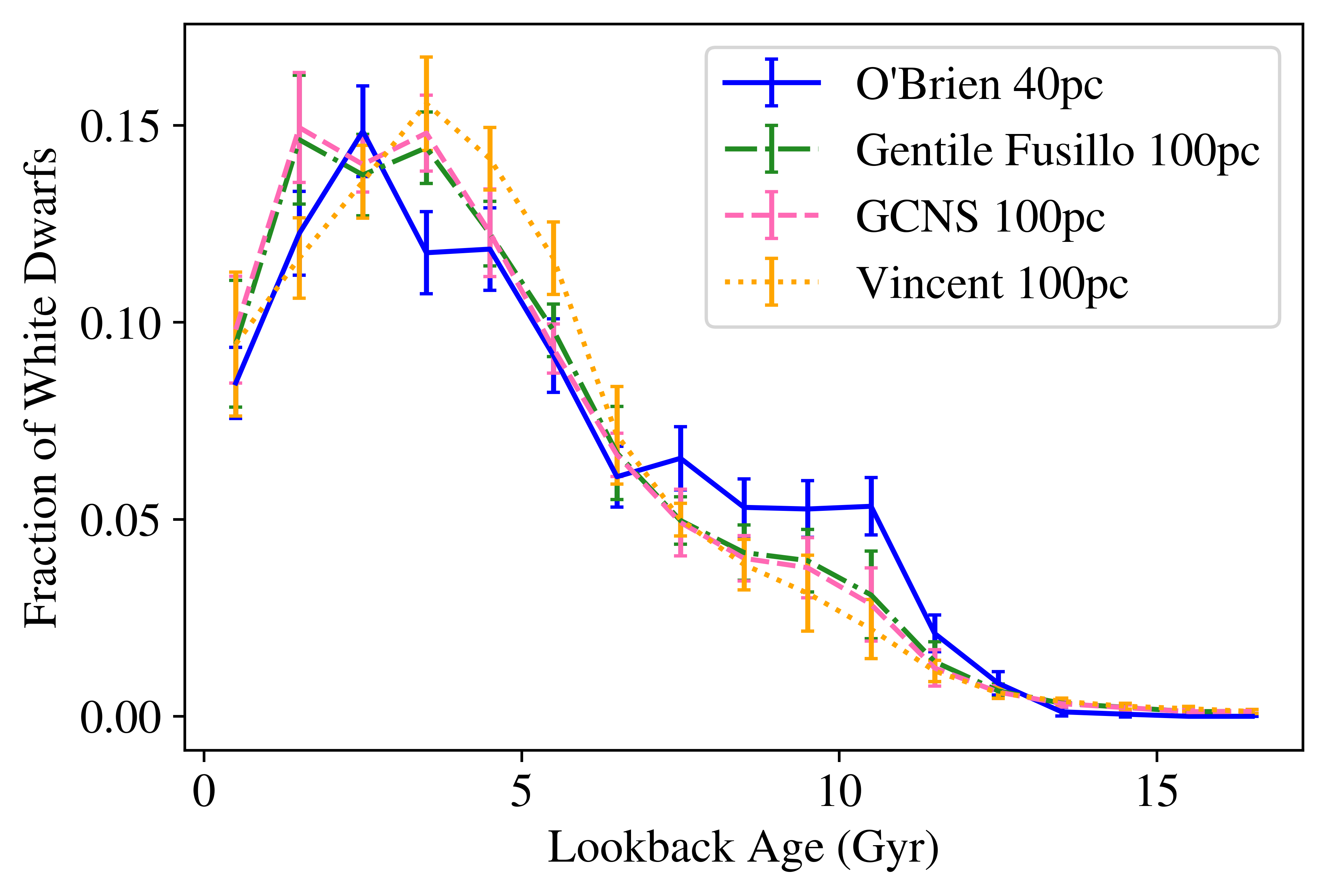}
    \caption{Star formation history derived from the direct white dwarf age method for the all-sky 100\,pc samples of \citet{GentileFusillo_2021,GCNS_2021,Vincent_2024} and the 40\,pc sample \citep{OBrien_2024,Roberts_2025}. Vertical errors are Poisson errors and the average spread of bin values across the 100 runs added in quadrature when systematic uncertainties and \textit{Gaia} observational errors are applied}
    \label{fig:100vs40pc_DA}
\end{figure}

In Fig.\,\ref{fig:100vs40pc_DA}, the directly calculated SFH of the Gentile Fusillo et al 100\,pc sample can be seen compared to that of the 40\,pc sample, and the other all-sky 100\,pc samples considered in this work. All of these derived star formation histories have been corrected for the missing main-sequence stars, and for stars that have formed within the volume under investigation but are not inside it within the present day due to the age-velocity dispersion relation (see \citealt{Roberts_2025} for details). The Gentile Fusillo et al direct age results show very good agreement with those of the 40\,pc sample: a very low rate of star formation at the beginning of the Galaxy's history, increasing to a higher rate of star formation within the last $\approx$5\,Gyr. The age of the population, marked by the oldest stars in the sample, sits in the 12-13\,Gyr bin but \textit{Gaia} data uncertainties, combined with the large sensitivity of total age to white dwarf mass, mean that this limit has large error bars, above 2\,Gyr either way.

\begin{figure}
	\includegraphics[width=0.9\columnwidth]{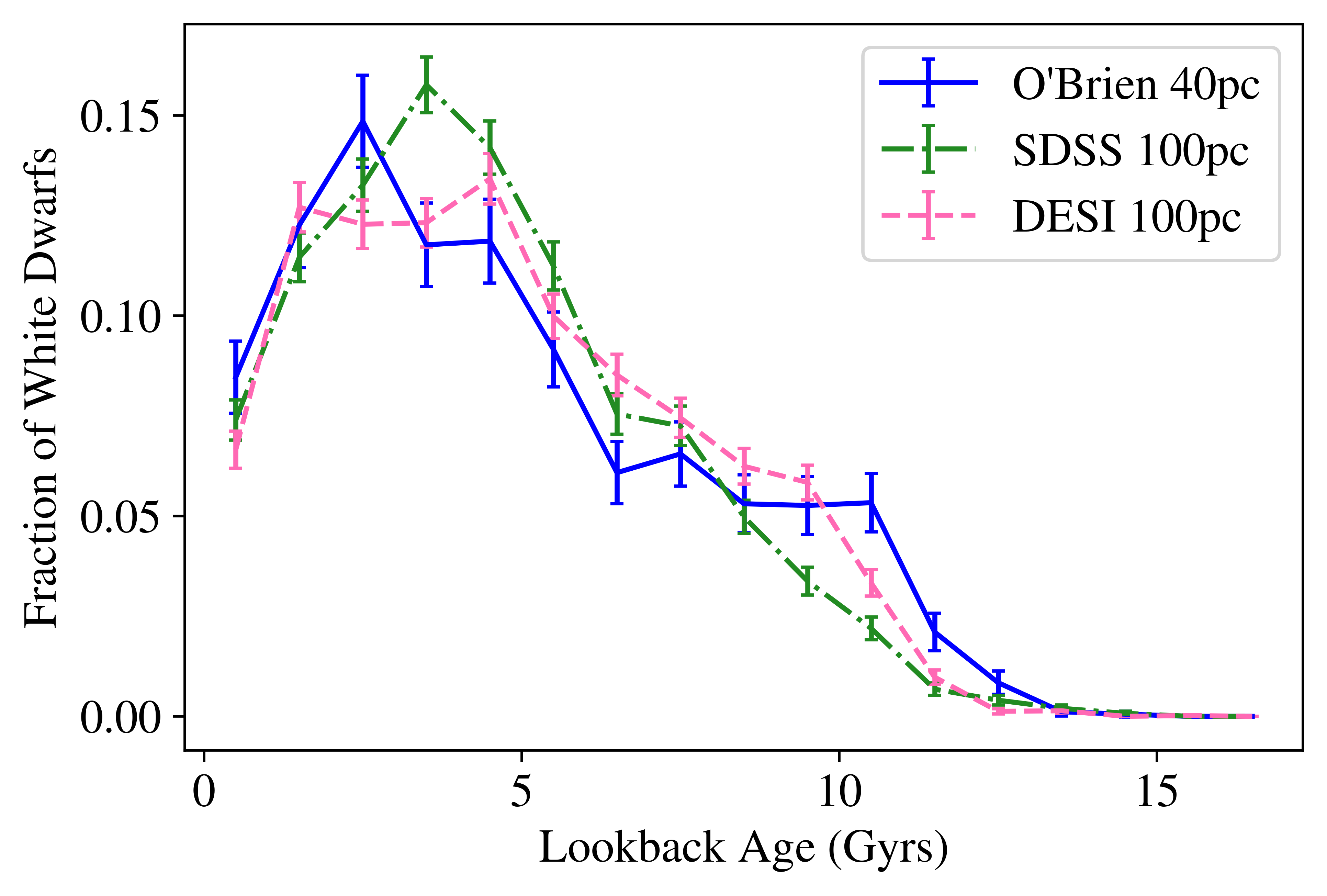}
    \caption{Similar to Fig.\,\ref{fig:100vs40pc_DA}, but for the SDSS \citep{Kilic_2025_100pc} and DESI \citep{DESI_2025} samples compared to 40\,pc.}
    \label{fig:100vs40pc_DA_partsky}
\end{figure}

The direct age result for the GCNS sample shown in Fig.\,\ref{fig:100vs40pc_DA} follows the line of the Gentile Fusillo et al sample almost exactly. 
The GSPC-WD sample does show some statistically significant differences compared to the other two samples: the peak of star formation occurs slightly earlier in the Galaxy's history, and the star formation rate appears to increase much more steeply in the 5-10\,Gyr range. This is likely also due to the colour (and therefore age) dependent biases discussed in the main text. With fewer faint white dwarfs in the sample, the normalised distribution of age is marginally skewed compared to the other, less biased samples. The SDSS sample in Fig.\,\ref{fig:100vs40pc_DA_partsky} clearly shows a similar behaviour, while the DESI sample is visually between the all-sky Gentile Fusillo et al result and the part-sky SDSS result.

\bsp	
\label{lastpage}
\end{document}